\makeatletter

\font\tenmsx=msxm10
\font\sevenmsx=msxm7
\font\fivemsx=msxm5
\font\tenmsy=msym10
\font\sevenmsy=msym7
\font\fivemsy=msym5
\newfam\msxfam
\newfam\msyfam
\textfont\msxfam=\tenmsx  \scriptfont\msxfam=\sevenmsx
  \scriptscriptfont\msxfam=\fivemsx
\textfont\msyfam=\tenmsy  \scriptfont\msyfam=\sevenmsy
  \scriptscriptfont\msyfam=\fivemsy

\def\hexnumber@#1{\ifnum#1<10 \number#1\else
 \ifnum#1=10 A\else\ifnum#1=11 B\else\ifnum#1=12 C\else
 \ifnum#1=13 D\else\ifnum#1=14 E\else\ifnum#1=15 F\fi\fi\fi\fi\fi\fi\fi}

\def\msx@{\hexnumber@\msxfam}
\def\msy@{\hexnumber@\msyfam}
\mathchardef\boxdot="2\msx@00
\mathchardef\boxplus="2\msx@01
\mathchardef\boxtimes="2\msx@02
\mathchardef\square="0\msx@03
\mathchardef\blacksquare="0\msx@04
\mathchardef\centerdot="2\msx@05
\mathchardef\lozenge="0\msx@06
\mathchardef\blacklozenge="0\msx@07
\mathchardef\circlearrowright="3\msx@08
\mathchardef\circlearrowleft="3\msx@09
\mathchardef\rightleftharpoons="3\msx@0A
\mathchardef\leftrightharpoons="3\msx@0B
\mathchardef\boxminus="2\msx@0C
\mathchardef\Vdash="3\msx@0D
\mathchardef\Vvdash="3\msx@0E
\mathchardef\vDash="3\msx@0F
\mathchardef\twoheadrightarrow="3\msx@10
\mathchardef\twoheadleftarrow="3\msx@11
\mathchardef\leftleftarrows="3\msx@12
\mathchardef\rightrightarrows="3\msx@13
\mathchardef\upuparrows="3\msx@14
\mathchardef\downdownarrows="3\msx@15
\mathchardef\upharpoonright="3\msx@16

\mathchardef\downharpoonright="3\msx@17
\mathchardef\upharpoonleft="3\msx@18
\mathchardef\downharpoonleft="3\msx@19
\mathchardef\rightarrowtail="3\msx@1A
\mathchardef\leftarrowtail="3\msx@1B
\mathchardef\leftrightarrows="3\msx@1C
\mathchardef\rightleftarrows="3\msx@1D
\mathchardef\Lsh="3\msx@1E
\mathchardef\Rsh="3\msx@1F
\mathchardef\rightsquigarrow="3\msx@20
\mathchardef\leftrightsquigarrow="3\msx@21
\mathchardef\looparrowleft="3\msx@22
\mathchardef\looparrowright="3\msx@23
\mathchardef\circeq="3\msx@24
\mathchardef\succsim="3\msx@25
\mathchardef\gtrsim="3\msx@26
\mathchardef\gtrapprox="3\msx@27
\mathchardef\multimap="3\msx@28
\mathchardef\therefore="3\msx@29
\mathchardef\because="3\msx@2A
\mathchardef\doteqdot="3\msx@2B

\mathchardef\triangleq="3\msx@2C
\mathchardef\precsim="3\msx@2D
\mathchardef\lesssim="3\msx@2E
\mathchardef\lessapprox="3\msx@2F
\mathchardef\eqslantless="3\msx@30
\mathchardef\eqslantgtr="3\msx@31
\mathchardef\curlyeqprec="3\msx@32
\mathchardef\curlyeqsucc="3\msx@33
\mathchardef\preccurlyeq="3\msx@34
\mathchardef\leqq="3\msx@35
\mathchardef\leqslant="3\msx@36
\mathchardef\lessgtr="3\msx@37
\mathchardef\backprime="0\msx@38
\mathchardef\risingdotseq="3\msx@3A
\mathchardef\fallingdotseq="3\msx@3B
\mathchardef\succcurlyeq="3\msx@3C
\mathchardef\geqq="3\msx@3D
\mathchardef\geqslant="3\msx@3E
\mathchardef\gtrless="3\msx@3F
\mathchardef\sqsubset="3\msx@40
\mathchardef\sqsupset="3\msx@41
\mathchardef\trianglerighteq="3\msx@44
\mathchardef\trianglelefteq="3\msx@45
\mathchardef\bigstar="0\msx@46
\mathchardef\between="3\msx@47
\mathchardef\blacktriangledown="0\msx@48
\mathchardef\blacktriangleright="3\msx@49
\mathchardef\blacktriangleleft="3\msx@4A
\mathchardef\blacktriangle="0\msx@4E
\mathchardef\triangledown="0\msx@4F
\mathchardef\eqcirc="3\msx@50
\mathchardef\lesseqgtr="3\msx@51
\mathchardef\gtreqless="3\msx@52
\mathchardef\lesseqqgtr="3\msx@53
\mathchardef\gtreqqless="3\msx@54
\mathchardef\Rrightarrow="3\msx@56
\mathchardef\Lleftarrow="3\msx@57
\mathchardef\veebar="2\msx@59
\mathchardef\barwedge="2\msx@5A
\mathchardef\doublebarwedge="2\msx@5B
\mathchardef\angle="0\msx@5C
\mathchardef\measuredangle="0\msx@5D
\mathchardef\sphericalangle="0\msx@5E
\mathchardef\varpropto="3\msx@5F
\mathchardef\smallsmile="3\msx@60
\mathchardef\smallfrown="3\msx@61
\mathchardef\Subset="3\msx@62
\mathchardef\Supset="3\msx@63
\mathchardef\Cup="2\msx@64

\mathchardef\Cap="2\msx@65

\mathchardef\curlywedge="2\msx@66
\mathchardef\curlyvee="2\msx@67
\mathchardef\leftthreetimes="2\msx@68
\mathchardef\rightthreetimes="2\msx@69
\mathchardef\subseteqq="3\msx@6A
\mathchardef\supseteqq="3\msx@6B
\mathchardef\bumpeq="3\msx@6C
\mathchardef\Bumpeq="3\msx@6D
\mathchardef\lll="3\msx@6E

\mathchardef\ggg="3\msx@6F

\mathchardef\circledS="0\msx@73
\mathchardef\pitchfork="3\msx@74
\mathchardef\dotplus="2\msx@75
\mathchardef\backsim="3\msx@76
\mathchardef\backsimeq="3\msx@77
\mathchardef\complement="0\msx@7B
\mathchardef\intercal="2\msx@7C
\mathchardef\circledcirc="2\msx@7D
\mathchardef\circledast="2\msx@7E
\mathchardef\circleddash="2\msx@7F
\def\ulcorner{\delimiter"4\msx@70\msx@70 }
\def\urcorner{\delimiter"5\msx@71\msx@71 }
\def\llcorner{\delimiter"4\msx@78\msx@78 }
\def\lrcorner{\delimiter"5\msx@79\msx@79 }
\def\yen{\mathhexbox\msx@55 }
\def\checkmark{\mathhexbox\msx@58 }
\def\circledR{\mathhexbox\msx@72 }
\def\maltese{\mathhexbox\msx@7A }
\mathchardef\lvertneqq="3\msy@00
\mathchardef\gvertneqq="3\msy@01
\mathchardef\nleq="3\msy@02
\mathchardef\ngeq="3\msy@03
\mathchardef\nless="3\msy@04
\mathchardef\ngtr="3\msy@05
\mathchardef\nprec="3\msy@06
\mathchardef\nsucc="3\msy@07
\mathchardef\lneqq="3\msy@08
\mathchardef\gneqq="3\msy@09
\mathchardef\nleqslant="3\msy@0A
\mathchardef\ngeqslant="3\msy@0B
\mathchardef\lneq="3\msy@0C
\mathchardef\gneq="3\msy@0D
\mathchardef\npreceq="3\msy@0E
\mathchardef\nsucceq="3\msy@0F
\mathchardef\precnsim="3\msy@10
\mathchardef\succnsim="3\msy@11
\mathchardef\lnsim="3\msy@12
\mathchardef\gnsim="3\msy@13
\mathchardef\nleqq="3\msy@14
\mathchardef\ngeqq="3\msy@15
\mathchardef\precneqq="3\msy@16
\mathchardef\succneqq="3\msy@17
\mathchardef\precnapprox="3\msy@18
\mathchardef\succnapprox="3\msy@19
\mathchardef\lnapprox="3\msy@1A
\mathchardef\gnapprox="3\msy@1B
\mathchardef\nsim="3\msy@1C
\mathchardef\napprox="3\msy@1D
\mathchardef\nsubseteqq="3\msy@22
\mathchardef\nsupseteqq="3\msy@23
\mathchardef\subsetneqq="3\msy@24
\mathchardef\supsetneqq="3\msy@25
\mathchardef\subsetneq="3\msy@28
\mathchardef\supsetneq="3\msy@29
\mathchardef\nsubseteq="3\msy@2A
\mathchardef\nsupseteq="3\msy@2B
\mathchardef\nparallel="3\msy@2C
\mathchardef\nmid="3\msy@2D
\mathchardef\nshortmid="3\msy@2E
\mathchardef\nshortparallel="3\msy@2F
\mathchardef\nvdash="3\msy@30
\mathchardef\nVdash="3\msy@31
\mathchardef\nvDash="3\msy@32
\mathchardef\nVDash="3\msy@33
\mathchardef\ntrianglerighteq="3\msy@34
\mathchardef\ntrianglelefteq="3\msy@35
\mathchardef\ntriangleleft="3\msy@36
\mathchardef\ntriangleright="3\msy@37
\mathchardef\nleftarrow="3\msy@38
\mathchardef\nrightarrow="3\msy@39
\mathchardef\nLeftarrow="3\msy@3A
\mathchardef\nRightarrow="3\msy@3B
\mathchardef\nLeftrightarrow="3\msy@3C
\mathchardef\nleftrightarrow="3\msy@3D
\mathchardef\divideontimes="2\msy@3E
\mathchardef\varnothing="0\msy@3F
\mathchardef\nexists="0\msy@40
\mathchardef\mho="0\msy@66
\mathchardef\thorn="0\msy@67
\mathchardef\beth="0\msy@69
\mathchardef\gimel="0\msy@6A
\mathchardef\daleth="0\msy@6B
\mathchardef\lessdot="3\msy@6C
\mathchardef\gtrdot="3\msy@6D
\mathchardef\ltimes="2\msy@6E
\mathchardef\rtimes="2\msy@6F
\mathchardef\shortmid="3\msy@70
\mathchardef\shortparallel="3\msy@71
\mathchardef\smallsetminus="2\msy@72
\mathchardef\thicksim="3\msy@73
\mathchardef\thickapprox="3\msy@74
\mathchardef\approxeq="3\msy@75
\mathchardef\succapprox="3\msy@76
\mathchardef\precapprox="3\msy@77
\mathchardef\curvearrowleft="3\msy@78
\mathchardef\curvearrowright="3\msy@79
\mathchardef\digamma="0\msy@7A
\mathchardef\varkappa="0\msy@7B
\mathchardef\hslash="0\msy@7D
\mathchardef\hbar="0\msy@7E
\mathchardef\backepsilon="3\msy@7F
\def\Bbb{\ifmmode\let\next\Bbb@\else
 \def\next{\errmessage{Use \string\Bbb\space only in math mode}}\fi\next}
\def\Bbb@#1{{\Bbb@@{#1}}}
\def\Bbb@@#1{\fam\msyfam#1}

\def\inv{^{\raise.15ex\hbox{${
  \scriptscriptstyle -}$}\kern-.05em 1}}

\def\Dsl{\,\raise.15ex\hbox{$/$}\mkern-13.5mu D}
\def\dsl{\raise.15ex\hbox{$/$}\kern-.57em\hbox{$\partial$}}

\def\lspace{\ifx\answ\bigans{}\else\qquad\fi}
\def\del{\partial}
 \def\CC{\hbox{{$\cal C$}}}
 
\def\CL{\hbox{{$\cal L$}}}

\def\CR{\hbox{{$\cal R$}}}

\def\CQ{\hbox{{$\cal Q$}}}  
\def\CO{\hbox{{$\cal O$}}} 
\def\CX{\hbox{{$\cal X$}}}

\def\lform{\hbox{$\sqcup$}\llap{\hbox{$\sqcap$}}}
\def\darr#1{\raise1.5ex\hbox{$\leftrightarrow$}
\mkern-16.5mu #1}

\def\h{{{1\over2}}}

\def\INT{{\textstyle \int\kern-.642em\int}}

\def\C{{\Bbb C}}

\def\eps{{\epsilon}}

\def\trace{{\rm Tr\, }}

\def\ant{{{\scriptstyle S}}}

\def\tens{\mathop{\otimes}}

\def\la{{\triangleright}}\def\ra{{\triangleleft}}

\def\span{{\rm span}}

\def\Ad{{\rm Ad}}

\def\ev{{\rm ev}}
\def\coev{{\rm coev}}

\def\id{{\rm id}}

\def\nquad{{\!\!\!\!\!\!}}
\def\nqquad{\nquad\nquad}
\def\eqn#1#2{\begin{equation}#2\label{#1}\end{equation}}

\def\o{{}_{(1)}}\def\t{{}_{(2)}}\def\th{{}_{(3)}}

\def\und#1{{\underline {#1}}}

\def\uo{{{}^{(1)}}}\def\ut{{{}^{(2)}}}

\def\text#1{\mbox{\rm #1}}
\def\note#1{}

\def\blacksquare{{\lform}}
\def\frac#1#2{{{#1\over#2}}}

\def\proof{\goodbreak\noindent{\bf Proof\quad}}

\def\endproof{{\ $\lform$}\bigskip }

\def\align#1{\begin{eqnarray*}#1\end{eqnarray*}}

\def\und#1{{\underline{#1}}}

\def\vect{{\bf t}}
\def\vecu{{\bf u}}

\def\<{\langle}
\def\>{\rangle}

\makeatother

\documentstyle[11pt,epsf]{article}

\def\thebibliography#1{\section*{REFERENCES}\list
 {[\arabic{enumi}]}{\settowidth\labelwidth{[#1]}\leftmargin\labelwidth
 \advance\leftmargin\labelsep
 \usecounter{enumi}}
 \def\newblock{\hskip .11em plus .33em minus -.07em}
 \sloppy
 \sfcode`\.=1000\relax}

\newtheorem{lemma}{Lemma}[section]
\newtheorem{propos}[lemma]{Proposition}
\newtheorem{example}[lemma]{Example}

\newtheorem{corol}[lemma]{Corollary}
\newtheorem{defin}[lemma]{Definition}

\begin{document}\baselineskip 25pt

{\ }\hskip 4.7in DAMTP/93-4 
\vspace{.5in}

\begin{center} {\Large QUANTUM AND BRAIDED LIE ALGEBRAS}
\baselineskip 13pt{\ }\\
{\ }\\ S. Majid\footnote{SERC Fellow and Fellow of Pembroke College,
Cambridge}\\ {\ }\\
Department of Applied Mathematics\\
\& Theoretical Physics\\ University of Cambridge\\ Cambridge CB3 9EW, U.K.
\end{center}

\begin{center}
20 March 1993\end{center}
\vspace{10pt}
\begin{quote}\baselineskip 13pt
\noindent{\bf ABSTRACT} We introduce the notion of a braided Lie algebra
consisting of a finite-dimensional vector space $\CL$ equipped with a bracket
$[\ ,\ ]:\CL\tens\CL\to \CL$ and a Yang-Baxter operator $\Psi:\CL\tens\CL\to
\CL\tens\CL$ obeying some axioms. We show that such an object has an enveloping
braided-bialgebra $U(\CL)$. We show that every generic $R$-matrix leads to such
a braided Lie algebra with $[\ ,\ ]$ given by structure constants $c^{IJ}{}_K$
determined from $R$. In this case $U(\CL)=B(R)$ the braided matrices introduced
previously. We also introduce the basic theory of these braided Lie algebras,
including the natural right-regular action of a braided-Lie algebra $\CL$ by
braided vector fields, the braided-Killing form and the quadratic Casimir
associated to $\CL$. These constructions recover the relevant notions for
usual, colour and super-Lie algebras as special cases. In addition, the
standard quantum deformations $U_q(g)$ are understood as the enveloping
algebras of such underlying braided Lie algebras with $[\ ,\ ]$ on $\CL\subset
U_q(g)$ given by the quantum adjoint action.
\end{quote}

\baselineskip 13pt
\tableofcontents
\baselineskip 22pt

\section{Introduction}

Many authors have sought a description of $U_q(g)$ and other quantum groups as
generated, in some sense, by a finite-dimensional `quantum Lie algebra' via
some kind of enveloping algebra construction (just as $U(g)$ is the universal
enveloping algebra of the Lie algebra $g$). Such a notion would be useful since
one would only have to work with the finite-dimensional Lie algebra instead of
the whole quantum group. It is also important for geometrical applications
where we might be interested in quantum vector fields generated by the action
not of general elements of the quantum group but by the action of the `Lie
algebra' elements.

We are in a situation here where a new mathematical concept is needed: quantum
groups $U_q(g)$ have various interesting choices of generators but which ones
should we look at, and what axioms should they obey?
One idea would be to attempt to build this on $g$ itself but with some kind of
deformed bracket obeying some kind of new axioms. In \cite[Sec. 2]{Ma:skl} we
have initiated a different approach based on the subspace $\{l^+Sl^-\}\subset
U_q(g)$ (where $l^\pm$ are the FRT generators of $U_q(g)$\cite{FRT:lie}). This
subspace is already well-known to be useful for certain kinds of computations
and we introduced on it some kind of `quantum Lie bracket' $[\ ,\ ]$ based on
the quantum adjoint action and defined by structure constants $c^{IJ}{}_K$.
This is recalled briefly in Section~2. Our goal in the present paper is to
develop this further onto an axiomatic framework for this bracket.

The natural bracket here does not obey of course the Jacobi identities, but
rather we find that it obeys naturally some kind of `braided-Jacobi identity'.
In this notion, which we introduce, the Yang-Baxter operator associated to the
action of $U_q(g)$ in the adjoint representation plays a central role. Armed
with suitable identities we show quite generally that brackets obeying them
allow one to generate an entire enveloping algebra.

The problem of defining some kind of braided-Lie algebra has been an open one
for some time. The reason is that in a braided setting the Yang-Baxter operator
or braided-transposition $\Psi$ does not have square 1. As a result there is no
action of the symmetric group and no notion of the Jacobi identity as
$0=[\xi,[\eta,\zeta]]+$cyclic. If we do suppose that $\Psi^2=\id$ then we are
in the symmetric or unbraided situation as studied in
\cite{Gur:yan}\cite{Sch:gen} and elsewhere. In this situation everything goes
through just as in the case of usual or super-Lie algebras. Unfortunately, this
case is extremely similar to the usual or super case (because $\Psi$ basically
has eigenvalues $\pm 1$) so no really new phenomena are obtained. Moreover, it
is too restrictive to deal with quantum groups of interest, such as
$U_q(sl_2)$.

Our approach to the problem is the following. In a series of papers we have
introduced the notion of braided group, see
\cite{Ma:bra}\cite{Ma:exa}\cite{Ma:lin} and others. These are a generalization
of quantum groups in which the elements are allowed to have braid statistics.
This means that they live in a braided tensor category where the tensor product
$\tens$ is commutative only up to a braided-transposition $\Psi$. Most
importantly for us now, we introduced the notion of a braided-cocommutative
object of this type. Only such braided-cocommutative objects could be truly
expected to be some kind of enveloping algebra. Thus we know the object which
we wish to emerge as the enveloping algebra of some kind of braided Lie
algebra. By studying the properties of the braided-adjoint action of such
objects, we can then deduce the right properties of the braided-Lie algebra
itself. These braided groups and the necessary Jacobi-like properties of the
braided-adjoint action  form the topic of Section~3.

In Section~4 we take the properties of the braided-adjoint action formally as a
set of axioms for a braided-Lie algebra. Here we no longer assume that we are
given a braided group but rather our main theorem is to show that such
braided-Lie algebras indeed generate a braided group or semigroup. By the
latter we mean a bialgebra in a braided category without necessarily an
antipode. Our main example is developed in Section~5 where we see that the
quantum-Lie algebras of Section~2 fit naturally into this axiomatic framework.
The construction works for a general $R$-matrix and in this case $U(\CL)$
recovers the braided matrices $B(R)$ introduced in \cite{Ma:exa}. There they
were introduced as a braided version of a quantum function algebra (like
functions on $M_n$) but the same braided matrices arise as a braided enveloping
bialgebra. It is interesting that only after further quotienting by
determinant-type (and other relations) does one recover precisely $U_q(g)$ in
this way: the braided matrices seem to be a natural covering algebra of these
objects and yet have properties like an enveloping algebra. We have already
identified the braided matrices covering $U_q(sl_2)$ as a form of Sklyanin
algebra at degenerate parameter value\cite{Ma:skl}, which we understand now as
$U(\CL)$ where $\CL$ is a braided-deformation of $gl_2$.

In a different direction we note that the action of such braided-Lie algebras
should naturally be some kind of braided-vector field. We demonstrate this in
Section~6 where we compute the right-regular action of the generators on
$U_q(g)$. This was announced in \cite{Ma:introp} and
our goal here is to give the full details. These braided-vector fields are
characterised by a matrix-Leibniz rule
\[(ab)\overleftarrow{\del}{}^i{}_j= a\cdot\Psi(b\tens
\overleftarrow{\del}{}^i{}_k)\overleftarrow{\del}{}^k{}_j\]
and are the left-invariant (and bicovariant) `vector fields' generated by
right-translations of the braided-Lie algebra generators on the braided group.
This can be contrasted with other constructions for differential operators on
quantum groups. The main difference is that we abandon in our notion of
braided-Lie algebras and braided-vector fields a commitment to the usual linear
form of the Leibniz rule. This is tied to the linear coproduct $\Delta
\xi=\xi\tens 1+1\tens \xi$. In general for a quantum group there are few such
primitive elements. Instead, we work more generally and consider the notions as
subordinate to a choice of (braided) coproduct $\Delta$. Aside from the
standard linear one, the matrix coproduct then suggests this matrix-notion of
Lie algebras and vector fields. For expressions that reduce in the classical
limit to usual infinitesimals one need only work with
${\del}{}^i{}_j-\delta^i{}_j$.

Finally, in Section~7 we give another application where the notion of a natural
finite-dimensional Lie algebra object is useful, namely to the definition of
braided-Killing form. This is provided by the quantum or braided-trace in the
adjoint representation of $U(\CL)$ on $\CL$. One the generators
$u^i{}_j-\delta^i{}_j$ one recovers in the classical limit and for standard
$R$-matrices the usual Killing form. As an unusual phenomenon we find that the
Killing form is made non-degenerate on $gl_2=sl_2\oplus u(1)$ by the process of
braided $q$-deformation. Moreover, our constructions work for any bi-invertible
$R$-matrix and we give formulae for the braided-Killing form $g^{IJ}$ in  terms
of it. In the invertible case it can be used to raise and lower indices (i.e.
to identify $\CL$ and $\CL^*$) and is Ad-invariant and braided-symmetric in a
suitable sense. As an application of the braided-Killing form we
compute the corresponding quadratic Casimir
\[ C=u^Iu^J g_{IJ}\]
in this invertible case.

\section{Quantum Lie Algebras}

This section provides some motivation for the constructions of the paper from
the point of view of quantum groups. It is perfectly possible to proceed
directly to the braided version in the next section and return only for some
details needed for the examples in subsequent sections. Throughout the present
section some familiarity with quantum groups is assumed.
We work over a field $k$ or (with care) a commutative ring (the reader can keep
in mind $\C$ or $\C[[\hbar]]$) and use the usual notations and methods for a
quasitriangular Hopf algebra $(H,\Delta,\eps,S,\CR)$. Here $H$ is a unital
algebra, $\Delta:H\to H\tens H$ is the coproduct, $\eps:H\to k$ the counit,
$S:H\to H$ the antipode and (for a strict quantum group) $\CR\in H\tens H$ is
the quasitriangular structure or so-called `universal R-matrix'. It obeys the
axioms of Drinfeld\cite{Dri},
\eqn{quasitr}{ (\Delta\tens\id)(\CR)=\CR_{13}\CR_{23},\quad
(\id\tens\Delta)(\CR)=\CR_{13}\CR_{12},\quad \sum h\t\tens h\o=\CR(\Delta\,
h)\CR^{-1}.}
For an introduction one can see \cite{Ma:qua}. Here and below we use the
Sweedler notation\cite{Swe:hop} $\Delta h=\sum h\o\tens h\t$ for the coproduct.

The problem which we consider is the following. It is well-known that the
standard quantum-groups $\CO_q(G)$ of function algebra type can be obtained by
an $R$-matrix method as
quotients of the quantum matrices $A(R)$ by determinant and other
relations\cite{FRT:lie}. On the other hand, the known treatments of the quantum
enveloping algebras to which these are dual, are quite a bit different. There
is the approach of Drinfeld and Jimbo in terms of the
roots\cite{Dri}\cite{Jim:dif} and an approach in \cite{FRT:lie} with twice as
many generators $l^\pm$ which have to be cut down somehow (usually by means of
some imaginative ansatz). These $l^\pm$ are roughly speaking the matrix
elements of the fundamental and conjugate-fundamental representation of $A(R)$.
Here we recall a somewhat different approach based on the quantum Killing form
and a single braided-matrix of generators $\vecu=(u^i{}_j)$ and developed in
\cite{Ma:skl}.

Just as Lie algebras like $sl_n$ can be defined both via root systems and via
matrices, so we give in this way a matrix approach to the standard quantum
enveloping algebras. At the same time a remarkable correspondence principle or
self-duality emerges between $\CO_q(G)$ as a quotient of quantum matrices
$A(R)$ and $U_q(g)$ as a corresponding quotient of the braided matrices $B(R)$.
In the present section we shall try to give a self-contained picture of this
using well-known quantum group formulae without too much direct dependence on
the theory of braided groups.

Let $R$ in  $M_n\tens M_n$ be an invertible matrix solution of the quantum
Yang-Baxter equations (QYBE) $R_{12}R_{13}R_{23}=R_{23}R_{13}R_{12}$. We recall
that $A(R)$ denotes the matrix bialgebra with generators $\vect=(t^i{}_j)$ and
relations $R\vect_1\vect_2=\vect_2\vect_1 R$ in the usual compact notation
(where the numerical suffices refer to the position in a matrix tensor
product). We recall also that $B(R)$ denotes the quadratic algebra with $n^2$
generators $\{ u^i{}_j\}$ and $1$, and relations
\eqn{B(R)}{R^k{}_a{}^i{}_b u^b{}_c R^c{}_j{}^a{}_d u^d{}_l=u^k{}_a
R^a{}_b{}^i{}_c u^c{}_d R^d{}_j{}^b{}_l\quad {\rm i.e.}\quad
R_{21}\vecu_1R_{12}\vecu_2= \vecu_2 R_{21} \vecu_1 R_{12}.}
These relations have been known for some time to be convenient for describing
$U_q(g)$ but they have been studied formally as a quadratic algebra for the
first time in \cite{Ma:exa}. We will come to the braided aspect\cite{Ma:exa} in
Section~5. For now we just work with $B(R)$ as a quadratic algebra.

\begin{propos} The algebra $B(R)$ is dual to $A(R)$ in the following sense. Let
$(H,\CR)$ be a quasitriangular bialgebra which is dually paired by $<\ ,\ >$
with $A(R)$ such that $<\vect_1\tens\vect_2,\CR>=R$. Let
\[  l=(\vect\tens\id)(Q),\qquad Q=\CR_{21}\CR_{12}.\]
Here $l^i{}_j$ are elements of $H$. Then there is an algebra map $B(R)\to H$
such that $l$ is the image of $\vecu$, i.e. $H$ is a realization of $B(R)$.
\end{propos}
\proof This is motivated by ideas for $U_q(g)$ implicit in the literature, see
\cite{ResSem:mat}\cite{ResSem:cen}. The new part in our approach however, is to
formulate the result at the level of bialgebras. Both $A(R)$ and $B(R)$ are
quadratic algebras and no antipode is needed. This approach arises out of the
transmutation theory of braided groups that related $A(R)$ to $B(R)$ in
\cite{Ma:eul}\cite{Ma:lin}, where we show the useful identity
\eqn{trans-mat}{l_1 R_{12}l_2=R_{12}(\vect_1\vect_2\tens\id)(Q).}
To be self-contained we can also give a direct proof of this easily enough as
\align{l_1 R_{12}l_2 &=& \sum R_{21}
<\vect_1,Q\uo><\vect_1\tens\vect_2,\CR><\vect_2,Q'\uo>Q\ut Q'\ut\\
&=&\sum <\vect_1,\CR\ut\CR'\uo\CR''''\uo><\vect_2,\CR''''\ut\CR''\ut\CR'''\uo>
\CR\uo\CR'\ut\CR''\uo\CR'''\ut\\
&=& \sum
\CR\uo\CR''\uo\CR'\ut\CR'''\ut<\vect_1,\CR\ut
\CR''''\uo\CR'\uo><\vect_2,\CR''\ut\CR''''\ut\CR'''\uo>\\
&=& \sum
<\vect_1,\CR\ut\t\CR''''\uo\CR'\uo\o><\vect_2,\CR\ut\o
\CR''''\ut\CR'\ut>\CR\uo\CR'\ut\\
&=& \sum
<\vect_1,\CR''''\uo\CR\ut\o\CR'\uo\o><\vect_2,
\CR''''\ut\CR\ut\t\CR'\ut>\CR\uo\CR'\ut\\
&=& \sum
R_{12}<\vect_1\vect_2,\CR\ut><\vect_1\vect_2,
\CR'\uo>\CR\uo\CR'\ut=R_{12}<\vect_1\vect_2\tens\id,Q\uo>Q\ut}
where $\CR',\CR''$ etc denote further copies of $\CR=\sum\CR\uo\tens\CR\ut$.
For the second equality we recognised the matrix form of the coproduct of the
$\vect$ as paired to multiplication in $H$. For the third equality we used the
QYBE for $\CR$. For the fourth and fifth we used the axioms (\ref{quasitr})
directly. We then wrote the expressions as products in $A(R)$ for the sixth
equality and recognized the result.

By permuting the matrix position labels we have equally well
$l_2R_{21}l_1=R_{21}(\vect_2\vect_1\tens\id)(Q)$. Hence
\[R_{21}l_1R_{12}l_2=R_{21}R_{12}(\vect_1\vect_2
\tens\id)(Q)=(\vect_1\vect_2\tens\id)(Q)R_{21}R_{12}=l_2R_{21}l_1 R_{12}\]
using the relations $R\vect_1\vect_2=\vect_2\vect_1R$ repeatedly. \endproof

In this sense then, $B(R)$ is some kind of universal dual algebra to $A(R)$.
Just as $A(R)$ has to be cut down by determinant and other relations to obtain
an honest Hopf algebra, likewise if $H$ is a Hopf algebra then $B(R)$ is
generally a little too big to coincide with $H$: it too has to be cut down by
additional relations. Note that in this case where $H$ is a Hopf algebra the
elementary identity $\CR^{-1}=(S\tens\id)(\CR)$ means that $l=l^+Sl^-$ relating
this description of $H$ to the FRT approach in \cite{FRT:lie}. For the next
proposition we concentrate on those standard quantum groups $U_q(g)$ which can
be put in this FRT form (this includes the deformations of at least the
non-exceptional semisimple Lie algebras).

\begin{propos}\cite{Ma:skl} Let $H=U_q(g)$ be of FRT form\cite{FRT:lie} with
associated R-matrix $R$ and dually paired with $A(R)$. Then the map $B(R)\to
U_q(g)$ has kernel given by `braided versions' of the determinant and other
relations associated to the Lie group $G$. Hence $U_q(g)$ can be identified as
$B(R)$ modulo such relations.
\end{propos}
\proof The argument in \cite{Ma:skl} is as a non-trivial corollary of the
process of transmutation\cite{Ma:eul}. This turns the matrix bialgebra $A(R)$
into the braided matrix $B(R)$ and also turns the quotient quantum groups
$\CO_q(G)$ into their braided versions $B_q(G)$. This is done in a categorical
way (by shifting categories) and transmutes at the same time all constructions
such as quantum planes etc on which these objects act. Hence (by these rather
general arguments) $B_q(G)$ is obtained in a braided version of the way that
$\CO_q(G)$ is obtained. On the other hand, there is also a braided version
$BU_q(g)$ of $U_q(g)$ coinciding as an algebra. Unlike the untransmuted theory
the quantum Killing form $Q:B_q(G)\to BU_q(g)$ is not just a linear map but a
map of braided Hopf algebras. For the standard  deformations of semisimple Lie
algebras it is even an isomorphism. This is the general reason for the
correspondence between $\CO_q(G)$ and $U_q(g)$ as quotients of matrices. For a
truly self-contained picture one can of course verify the proposition directly
by computing in detail the required quotients of $B(R)$. For example, for
$U_q(sl_2)$ on must divide $BSL_q(2)$ by the braided determinant
$ab-q^2cb=1$\cite{Ma:exa}. \endproof

Note that the additional relations needed to obtain a Hopf algebra like
$U_q(g)$ in this way from $B(R)$ are such that there exists a braided antipode
$\und S$ with $\vecu\und S \vecu=1=(\und S \vecu)\vecu$. This exhibits the
remarkable similarity with what is done to obtain a quantum group from $A(R)$,
but with one catch: the matrix coproduct $\und \Delta \vecu=\vecu\tens\vecu$
does not give a bialgebra in the usual sense (it is not an algebra homomorphism
to the usual tensor product). This explains why for a full appreciation of this
approach one must understand $B(R)$ correctly as a bialgebra with
braid-statistics\cite{Ma:exa}.

Even without such a full picture, Proposition~2.2 does however, provide a quick
way of computing $U_q(g)$ as well as the corresponding quantum enveloping
algebra in a general non-standard but factorizable case. Namely, compute the
quadratic algebra $B(R)$ and then impose further determinant-type and other
relations.  Factorizable means here by definition that the map from the
relevant dual of $H$ to $H$ given by evaluation against the first factor of
$Q=\CR_{21}\CR_{12}$ is a surjection\cite{ResSem:mat}. This ensures in
Proposition~2.1 that for such $H$ the $l$ are generators. Note also that the
existence of a quasitriangular Hopf algebra dually paired to $A(R)$ is not
possible for all $R$. A necessary condition is thet $R$ has a second inverse
$\widetilde{R}=<\vect_1\tens \vect_2,(\id\tens S)\CR>=((R^{t_2})^{-1})^{t_2}$
(where $t_2$ is transposition in the second matrix factor).

For the moment we can just note then that $U_q(g)$ and in general any
quasitriangular Hopf algebra dual to $A(R)$ has a subspace
\eqn{subspaceL}{ \CL=\span<l^i{}_j>\subset H}

\begin{propos}\cite{Ma:skl}

i) In the factorizable case\cite{ResSem:mat}, the subspace $\CL=\span<l^i{}_j>$
and $1$ generate all of $H$.
ii) The subspace  $\CL$ is stable under the quantum adjoint action of $H$ on
itself.

iii) The quantum adjoint action as a map $[\ ,\ ]:\CL\tens\CL\to \CL$ looks
explicitly like
\[ [l^I,l^J]=c^{IJ}{}_K l^K,\quad
c^{IJ}{}_K=\widetilde{R}{}^a{}_{i_1}{}^{j_0}{}_b R^{-1}{}^b{}_{k_0}{}^{i_0}{}_c
R^{k_1}{}_{n}{}^c{}_m R^{m}{}_a{}^{n}{}_{j_1}\]
where $l^I=l^{i_0}{}_{i_1}$ and $I=(i_0,i_1)$ is a multi-index notation
(running from $(1,1),\cdots,(n,n)$).

\end{propos}
\proof Again, the proof in \cite{Ma:skl} is based on the theory of
transmutation in \cite{Ma:eul}\cite{Ma:lin}. The linear space of $B(R)$ can be
identified with that of $A(R)$ with the generators $\vecu=\vect$ identified
(but not their products as we have seen above). Then (ii) is automatic because
$\vect$ transforms to a linear combination under the quantum coadjoint action,
hence so does $\vecu$ under the quantum adjoint action. To be self-contained we
can also give a direct proof using more familiar methods as follows.

(i) In the present setting this is (as we have mentioned) more or less by the
definition of factorizable. In our usage  this notion is subordinate to the
choice of a bialgebra or Hopf algebra dually paired with $H$ in the sense of
\cite{Ma:qua}. Here the choice is $A(R)$ or its quotients such as $\CO_q(G)$.

(ii) We use the form $l=l^+Sl^-$ valid in the Hopf algebra case and let
$\la=\Ad$ denote the quantum adjoint action of $H$ on itself given by $h\la
b=\sum h\o b S h\t$. We show that\cite{Ma:seq}
\eqn{lpm-Ad-l}{l^+_2\la l_1=R^{-1}l_1R,\qquad l^-_1\la l_2=Rl_2 R^{-1}}
using the definition of $\Ad$, elementary properties of the antipode and the
fact that $l^\pm$ obey the relations $R^{-1}l^\pm_1 l^\pm_2=l^\pm_2 l^\pm_1
R^{-1}$ and $R^{-1}l^-_1 l^+_2=l^+_2 l^-_1 R^{-1}$ as in \cite{FRT:lie} (These
relations are not tied to the standard $U_q(g)$ as in \cite{FRT:lie} if one
uses the general formulation as in \cite{Ma:qua}\cite{Ma:seq}). Thus
\align{l^+_2\la (l^+_1Sl^-_1)&&=l^+_2l^+_1Sl^-_1 Sl^+_2=l^+_2l^+_1 S(l^+_2
l^-_1)=l^+_2l^+_1R^{-1} S( Rl^+_2 l^-_1)\\
&&=R^{-1}
l^+_1l^+_2 S(l^-_1 l^+_2)R=R^{-1}l^+_1(l^+_2Sl^+_2)Sl^-_1R=R^{-1}l^+_1Sl^-_1
R\\
l^-_1\la (l^+_2Sl^-_2)&&=l^-_1l^+_2Sl^-_2 Sl^-_1=l^-_1l^+_2 S(l^-_1
l^-_2)=l^-_1l^+_2R S( R^{-1}l^-_1 l^-_2)\\
&&=R l^+_2l^-_1 S(l^-_2
l^-_1)R^{-1}=Rl^+_2(l^-_1Sl^-_1)Sl^-_2R^{-1}=Rl^+_2Sl^-_2 R^{-1}.}

(iii) We can also deduce from this the action of $Sl^\pm$ from
$l^+Sl^+=\id=(Sl^+)l^+$ etc (the identity matrix times the action of the
identity). In particular,
\eqn{Slm-Ad-l}{ (\ant l^-{}^i{}_j)\la l^k{}_l=\widetilde{R}{}^a{}_j{}^k{}_m
l^m{}_n R^i{}_a{}^n{}_l.}
where $\widetilde{R}$ obeys $\widetilde{R}^i{}_a{}^b{}_l
R^a{}_j{}^k{}_b=\delta^i{}_j\delta^k{}_l=R^i{}_a{}^b{}_l
\widetilde{R}^a{}_j{}^k{}_b.$
Combining this with (\ref{lpm-Ad-l}) we can compute $l^{i_0}{}_{i_1}\la
l^{j_0}{}_{j_1}=l^+{}^{i_0}{}_a\la((Sl^-{}^a{}_{i_1})
\la l^{j_0}{}_{j_1}$ to find the result stated.
\endproof

Thus $\CL$ is some kind of `quantum Lie algebra' for $H$ because it is a
finite-dimensional subspace that generates $H$ and at the same time is closed
under the quantum adjoint action, which provides a kind of `quantum Lie
bracket' $[\xi,\eta]=\xi\la\eta$. We have introduced this point of view in
\cite{Ma:skl} and pointed out that this bracket obeys a number of
Lie-algebra-like identities inherited from the standard properties of the
quantum adjoint action, such as

(L0) $\qquad [\xi,\eta]\in\CL\qquad {\rm for} \qquad\xi,\eta\in \CL$

(L1) $\qquad [\xi,[\eta,\zeta]]=\sum [[\xi\o,\eta],[\xi\t,\zeta]]$

(L$1'$) $\qquad [[\xi,\eta],\zeta]=\sum [\xi\o,[\eta,[ \ant\xi\t,\zeta]]]$

The second of these is just the statement that $\Ad$ is a covariant action of
the Hopf algebra on itself, while the third follows from the definition of
$\Ad$. We see here two problems with this approach. Firstly, these properties
(L1) and (L$1'$) cannot be taken as abstract properties of some kind of Lie
algebra structure on $\CL$ because they involve the coproduct and this does
{\em not} in general act on $\CL$ in a simple way (its just gives some subspace
of $H\tens H$). Secondly, they hold for any Hopf algebra and so do not express
the fact that quantum groups such as $U_q(g)$ are close to being cocommutative.
The usual Lie bracket has properties inherited from the fact that $U(g)$ is
cocommutative, and our quantum Lie algebra, to be convincing, should deform
some of these. For later reference,

\begin{propos} If $H$ is a cocommutative Hopf algebra then the usual Hopf
algebra adjoint action $[\ ,\ ]=\Ad$ obeys in addition to the identities above,
the identities

(L2) $\qquad\sum \xi\t\tens [\xi\o,\eta]=\sum \xi\o\tens [\xi\t,\eta]$

(L3) $\qquad\sum [\xi,\eta]\o\tens [\xi,\eta]\t=\sum [\xi\o,\eta\o]\tens
[\xi\t,\eta\t]$

for all $\xi,\eta$ in $H$.
\end{propos}
\proof (L2) needs no comment except to say that we have written the
cocommutativity in this way because later
we shall adopt something like this without assuming that the Hopf algebra is
completely cocommutative. This weak notion of cocommutativity (as relative to
something on which the Hopf algebra acts) is useful in other contexts also. For
(L3) we have $\Delta[\xi,\eta]=\sum \xi\o\o \eta\o S\xi\t\t\tens \xi\t\o\eta\t
S\xi\t\o$ using that $S$ is an anticoalgebra map. In the cocommutative case the
numbering of the suffices does not matter so we have at once the right hand
side of (L3). For the record we give here also the proof of (L1). This holds
for any Hopf algebra and reads
\align{\sum [[\xi\o,\eta],[\xi\t,\zeta]]\nqquad&&=\sum (\xi\o\o \eta
S\xi\o\t)\o (\xi\t\o\zeta S\xi\t\t) S(\xi\o\o \eta S\xi\o\t)\t\\
&&=\sum \xi\o\eta\o (S\xi_{(4)})\xi_{(5)}\zeta (S\xi_{(6)})
(S^2\xi\th)(S\eta\t)(S\xi\t)\\
&&=\sum \xi\o\eta\o \zeta (S\xi_{(4)}) (S^2\xi\th)(S\eta\t)(S\xi\t)=\xi\o\eta\o
\zeta (S\eta\t)(S\xi\t)=[\xi,[\eta,\zeta]]}
expanding out the definitions, the properties of the antipode and the Sweedler
notation \cite{Swe:hop} to renumber the suffices to base 10 (keeping the
order). The third and fourth equalities then successively collapse using the
axioms of an antipode. \endproof

Another aspect of our matrix approach, which is not a problem but a convention
is that our chosen finite-dimensional subspace $\CL$ is a mixture of
`group-like' elements with coproduct $\Delta\xi\sim \xi\tens\xi$ and more usual
Lie-algebra-like elements where $\Delta\xi\sim \xi\tens 1+1\tens\xi$ (with a
suitable deformation). The latter are how off-diagonal elements of $l^i{}_j$
tend to behave, while the former are how diagonal elements tend to behave.
Another good convention is to take as `quantum Lie algebra' the subspace
\eqn{L-1}{ \CX=\span<\chi^i{}_j>\subset H,\quad
\chi^i{}_j=l^i{}_j-\delta^i{}_j}
This is a matter of taste and is entirely equivalent. The subspace is also
closed under $[\ ,\ ]=\Ad$ which now has structure constants
\eqn{liechi}{
[\chi^I,\chi^J]=[l^I,l^J]+\delta^I\delta^J-\delta^I\delta^J-\delta^I
l^J=(c^{IJ}{}_K-\delta^I\delta^J{}_K)\chi^K}
using the elementary properties of the quantum adjoint action (notably $l^I\la
1=\eps(l^I)=\delta^I$). Here $\delta^I=\delta^{i_0}_{i_1}$ and $\delta^J{}_K$
are Kronecker delta-functions. The last equality uses that
\eqn{cdelta}{c^I{}^J{}_K\delta^K=\delta^I\delta^J}
which follows at once from the expression for $c$ in the proposition. These
$\chi^I$ equally generate $H$ along with $1$ and have a better-behaved
semi-classical limit in the standard cases. This aspect of our approach has
been stressed in \cite{SWZ:bic}. It is also quite natural from the point of
view of bicovariant differential calculus as explained in \cite{Sud:dif}. We
note also that some combinations of the basis elements of $\CL$ or $\CX$ can
have trivial quantum Lie bracket and have to be decoupled if we want to have
the minimum number of generators.

Finally, to complete the picture of $B(R)$ as a some kind of dual of $A(R)$ we
have an elementary lemma which we will need later in Section~6.

\begin{lemma} The generators of $A(R)$ define matrix elements of the
representation $\rho$ of $B(R)$ given by
\[ \rho_2(\vecu_1)=<\vecu_1,\vect_2>=Q_{12},\quad Q=R_{21}R_{12}\]
\end{lemma}
\proof We have to show that this extends consistently to all of $B(R)$ as an
algebra representation,
\align{\rho_3(R_{21}\vecu_1R_{12}\vecu_2)&=&R_{21}
\rho_3(\vecu_1)R_{12}\rho_3(\vecu_2)\\
&=& R_{21} Q_{13} R_{12} Q_{23}= Q_{23} R_{21} Q_{13} R_{12}\\
&=& \rho_{3}(\vecu_2)R_{21} \rho_3(\vecu_1)R_{12}=\rho_3(\vecu_2 R_{21}\vecu_1
R_{12}).}
The middle equality follows from repeated use of the QYBE. Thus the extension
is consistent with the algebra relations of $B(R)$. \endproof

One can also see over $\C$ that if $R$ obeys a certain reality condition then
$B(R)$ is a $*$-algebra with $u^i{}_j{}^*=u^j{}_i$. We call this the hermitian
real form of the braided matrices $B(R)$. At the level of the standard $U_q(g)$
with real $q$ the corresponding $l^i{}_j{}^*=l^j{}_j$ recovers the standard
compact real form of the these Hopf algebras. These remarks confirm that the
braided-matrix approach to $U_q(g)$ is quite natural.

\section{Properties of the Braided-Adjoint Action}

In this section we recall some basic facts about Hopf algebras in braided
categories (braided Hopf-algebras) and
their braided adjoint action. It is these categorical constructions that lead
to the notion of braided-Lie algebras in the next section. The idea is that we
know what is a braided group\cite{Ma:bg} (or in physical terms a group-like
object with braid statistics\cite{Ma:bra}) and we just have to infinitesimalize
this notion.

One of the novel aspects of braided groups is that results fully analogous to
those familiar in algebra or group
theory are proven now using braid and knot diagrams. This is because we work in
a braided or quasitensor category.
This means $(\CC,\tens,\und 1,\Phi,\Psi)$ where $\CC$ is a category (a
collection of objects and allowed morphisms or maps between them), $\tens$ is a
tensor product between two objects, with $\und 1$ a unit object for the tensor
product. The isomorphisms $\Phi_{V,W,Z}:V\tens (W\tens Z)\to (V\tens W)\tens Z$
express associativity and say that we can forgot about brackets (all tensor
products can be put into a canonical form in a consistent way). Finally, there
is a braided-transposition or quasisymmetry $\Psi_{V,W}:V\tens W\to W\tens V$
saying that the tensor product is commutative up to this isomorphism. The
difference between this setting and the standard one for symmetric monoidal
categories\cite{Mac:cat} is that we do not suppose that
$\Psi_{W,V}\circ\Psi_{V,W}=\id$. Put another way, we distinguish
carefully between $\Psi_{V,W}$ and $(\Psi_{W,V})^{-1}$ which are both morphisms
$V\tens W\to W\tens V$ for any two objects. To avoid confusion a good notation
here is to write the morphisms not in the usual way as single arrows, but
downward as braid crossings,
\[ \epsfbox{psi.eps}\]
These braided-transpositions do however obey other obvious properties of usual
transposition such as $\Psi_{V\tens W,Z}=\Psi_{V,Z}\circ\Psi_{W,Z}$ and
similarly for $\Psi_{V,W\tens Z}$. These ensure that different sequences of
braided transpositions that connect two composite objects coincide if the
corresponding braids in the notation above coincide. Also, these isomorphisms
are functorial in that they are compatible with any morphisms between objects.
If we write any morphisms also pointing downwards as notes with input lines and
output lines, then the functoriality says that we can
pull such nodes through braid crossings much as beads on a string.

\begin{figure}
\vskip .3in
\epsfbox{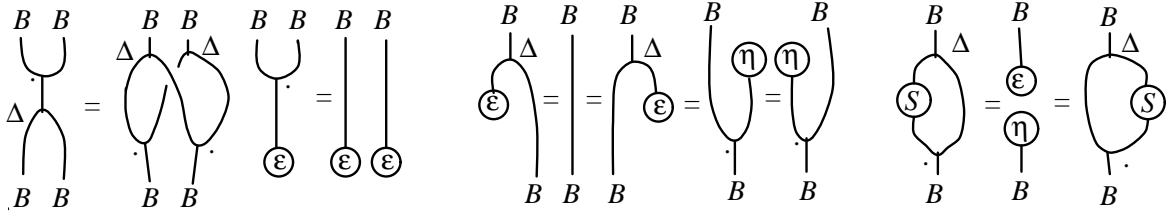}
\caption{Axioms of a Braided Hopf Algebra}
\end{figure}

This describes the diagrammatic notation that we shall use. For a formal
treatment of braided categories see \cite{JoyStr:bra} and for on introduction
to our methods see \cite[Sec. 3]{Ma:introp}. In this notation, the
axioms of a Hopf algebra in a braided category are recalled in Figure~1. They
are like a usual Hopf algebra $B$ except that the product, coproduct $\Delta$,
antipode $S$, unit $\eta$ and counit $\eps$ are all morphisms in the braided
category. In the diagrammatic notation we write the unit object as the empty
set. The first axiom shown (the bialgebra axiom) is the most important: it says
that the braided coproduct $B\to B\und\tens B$ is an algebra homomorphism where
$B\und\tens B$ is the braided tensor product algebra structure on $B\tens B$.
This is like a super-tensor product and involves transposition by $\Psi$. The
two factors $B$ in $B\und\tens B$ do not commute but instead enjoy braid
statistics given by $\Psi$.

The reader can keep in mind the trivially braided group (the ordinary group
Hopf algebra) $B=\C G$ with $\Delta g=g\tens g$ and $S g=g^{-1}$. The antipode
axiom says that if we split $g$ into $g,g$, apply $S$ to one factor and then
multiply up, we get something trivial. The diagrams on the right in Figure~1
just say this abstractly as morphisms.

It is remarkable that such objects defined in this way really behave like usual
groups or quantum groups. For example, the usual adjoint action of a group on
itself consists in taking $g, a$, splitting $g$ to give $g,g,a$, applying $S$
to give $g,g^{-1},a$, transposing  $g^{-1}$ past the $a$, and then multiplying
up. When written as diagrams or morphisms in our braided category, this is the
{\em braided adjoint action}. It is shown in the box in  Figure~2.

\begin{figure}
\vskip .3in
\epsfbox{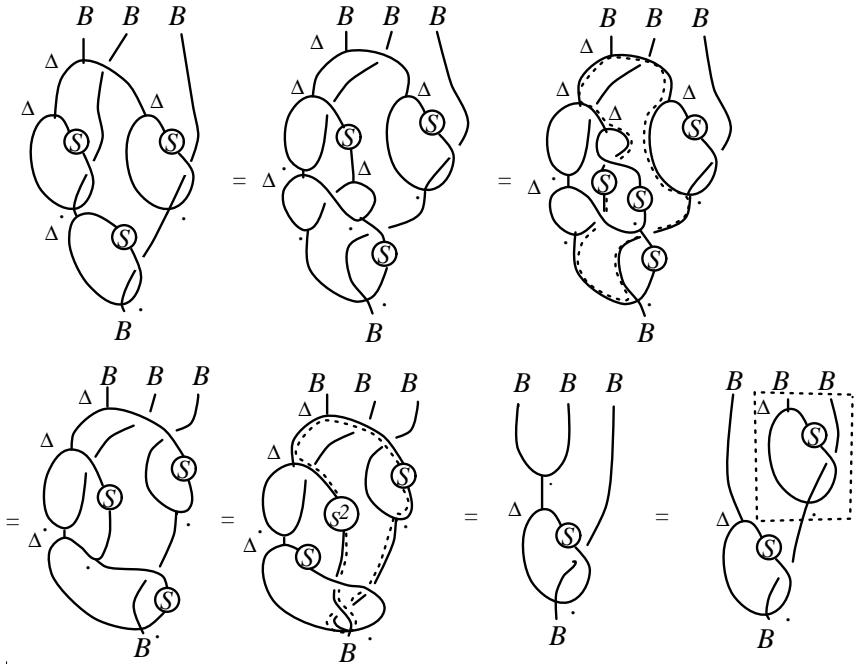}
\caption{Proof that the braided adjoint action obeys the braided Jacobi
identity}
\end{figure}

Figure~2 itself is the diagrammatic proof of the main result of this section.
It shows that applying the braided adjoint action twice as on the right in
Figure~2, is the same as the left hand expression. This consists in applying
the tensor product braided adjoint action of $B$ on $B\tens B$ and then
applying the adjoint action again to the result (all together three
applications of the braided adjoint action on the left in Figure~2). We call
this the {\em braided-Jacobi identity}. The proof reads as follows. Starting on
the left, use the bialgebra axiom that $\Delta$ is an algebra homomorphism to
expand the expression on the left. For the second equality we use the fact that
$S$ is a braided anti-coalgebra map  $\Delta\circ S= (S\tens
S)\Psi\circ\Delta$\cite{Ma:tra}. We then identify (the dotted line) a closed
loop which will after reorganization or the branches using associativity and
coassociativity cancel according to the antipode axioms in Figure~1. We make
this cancellation for the third equality. We then use that $S$ is a braided
anti-algebra map for the fourth equality and identify another loop. This
cancels in a similar way to the antipode loop giving the fifth equality. The
final equality is the easier fact already proven in \cite{Ma:lin} that the
braided adjoint action is indeed an action. We summarise this along with some
other known properties.

\begin{propos} Let $B$ be a Hopf algebra in a braided or quasitensor category
and let $\Ad=\cdot^2\circ(\id\tens\Psi_{B,B})(\id\tens
S\tens\id)(\Delta\tens\id)$ denote the braided adjoint action as above.
It is (a) an action of $B$ on itself  and (b) respects its own product (a
braided module algebra) as in Figure~3. Further (c) it obeys the braided Jacobi
identity. Finally, if $B$ is cocommutative with respect to $\Ad$ in the sense
of \cite{Ma:bra} (as shown) then (d) holds.
\end{propos}

\begin{figure}
\vskip .3in
\epsfbox{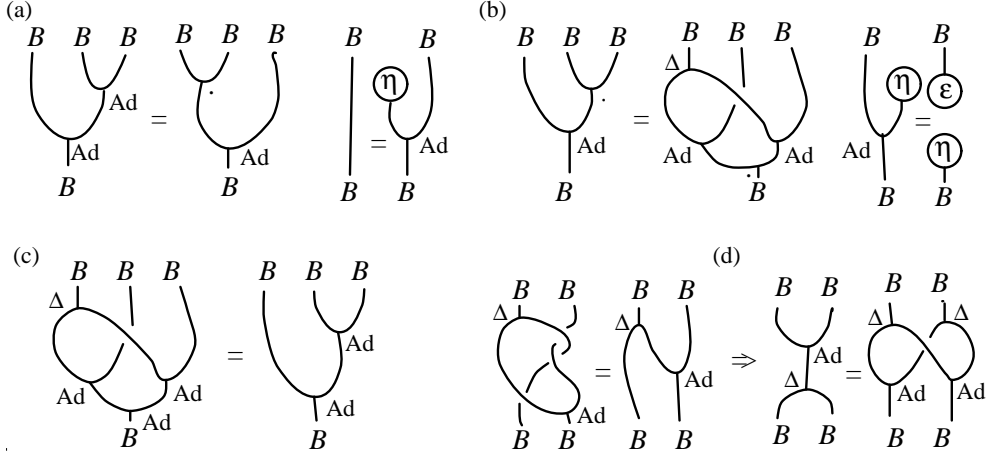}
\caption{Summary of Properties of the Braided Adjoint Action (a) an action (b)
a module-algebra under the action (c) the braided Jacobi identity (d)
compatibility with the coproduct implied by the assumption of
braided-cocommutativity with respect to $\Ad$}
\end{figure}

\proof We have spelled out the proofs of (a),(b) and (d) in \cite{Ma:lin} in a
dual form with comodules and coactions (for the braided adjoint coaction). We
ask the reader to turn the diagrammatic proofs for these in \cite{Ma:lin}
up-side down (a 180 rotation) and read them again. They read exactly as the
required proof for the $\Ad$ action. This is part of the self-duality of the
axioms of a Hopf algebra. For the new part (c) we have given the proof above.
\endproof

Note that for a usual non-cocommutative Hopf algebra the quantum adjoint action
does {\em not} respect the coproduct in the sense of (c) above. One needs a
cocommutativity condition. The idea in \cite{Ma:bra} was not to try to define
this intrinsically (the naive notion does not work well) but in a week form as
cocommutative with respect to a module. This is the form that we have used: we
suppose that $B$ is cocommutative in this weak sense. This corresponds to
directly assuming (L2) in Proposition~2.4. This is then enough to derive (d)
which corresponds to (L3) in that proposition.

To this extent then, the kind of Hopf algebras in braided categories that we
consider are truly like groups or enveloping algebras in the sense that they
are supposed  braided-cocommutative at least with respect to their own braided
adjoint action. This completes our review of the braided adjoint action and the
derivation of the identities that we will need in the next section. We will
take them as the defining properties of a braided-Lie algebra.

\section{Braided-Lie Algebras and their Enveloping Algebras}

We have seen that if we do have a braided group as in the last section then the
braided-Adjoint action obeys some
Lie-algebra like identities as in the second line in Figure~3. If the braided
group has some generating subobject
which is closed under $\Ad$ then these identities hold for it also. Motivated
by this, we are going to adopt these as abstract axioms for a braided Lie
algebra and prove a theorem in the converse direction. Thus every such braided
Lie algebra will have (at least in a category with direct sums) an enveloping
braided-bialgebra returning us to something like the the kind of braided group
we might have began with. One surprise will be that the enveloping algebra here
seems more naturally to be a bialgebra (in a braided sense) rather than a Hopf
algebra with antipode. Of course one can add further conditions to force a
braided-antipode but they do not appear to be very natural from the point of
view of the underlying braided Lie algebra.

\begin{defin} A braided Lie algebra is $(\CL,\Delta,\eps,[\ ,\ ])$ where $\CL$
is an object in a braided or quasitensor category, $\Delta:\CL\to \CL\tens\CL$
and $\eps:\CL\to\und 1$ are morphisms forming a coalgebra in the category, and
$[\ ,\ ]:\CL\tens\CL\to \CL$ is a morphism obeying the conditions
(L1),(L2),(L3) in Figure~4.
\end{defin}

\begin{figure}
\vskip .3in
\epsfbox{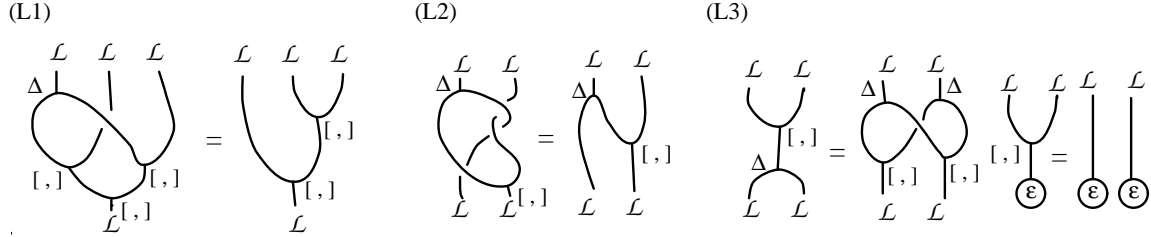}
\caption{Axioms of a Braided Lie Algebra (a) Braided-Jacobi identity axiom (b)
Cocommutativity axiom (c) Coalgebra compatibility axiom }
\end{figure}

The idea of introducing a coalgebra here is one of the novel aspects of the
approach. In the usual definition of a Lie algebra a coalgebra structure
$\Delta\xi=\xi\tens 1+1\tens\xi$ and $\eps\xi=0$ is implicit. We do not want to
be tied to a specific form such as this and hence bring the implicit $\Delta$
to the foreground as part of the axiomatic structure. The only requirements of
a coalgebra are
\eqn{bracoalg}{(\Delta\tens\id)\circ\Delta=(\id\tens\Delta)\circ\Delta, \quad
(\eps\tens\id)\circ\Delta=\id=(\id\tens\eps)\circ\Delta}
as usual.

There is no bialgebra axiom here because after all $\CL$ is not being required
to have an associative product. It is typically some finite-dimensional vector
space. Instead axiom (L1) says that $\CL$ is being equipped with some kind of
Lie bracket $[\ ,\ ]$. This braided-Jacobi identity is a form of associativity.
If one imagines momentarily the usual linear form for $\Delta$ then the left
hand side of (L1) has two terms then we have something like the usual Jacobi
identity as discussed in Section~2. We of course do not suppose this (we do not
even suppose that $\CL$ has an element that can be called $1$). We do however
suppose that $\Delta$ is braided-cocommutative with respect to this Lie bracket
$[\ ,\ ]$ in the sense of (L2) and that $\Delta$ respects it in the sense of
(L3). This (L3) is a Lie form of the bialgebra condition in Figure~1.

\begin{propos} Let $(\CL,\Delta,\eps,[\ ,\ ])$ be a braided Lie algebra in an
Abelian braided tensor category (we suppose that we have direct sums with the
usual properties). Then there is a braided bialgebra  $U(\CL)$ in the sense of
Section~3, generated by $1$ and $\CL$ with relations as shown in Figure~5. We
call it the {\em universal enveloping algebra} of the braided Lie algebra.
\end{propos}
\begin{figure}
\vskip .3in
\[\epsfbox{envreln.eps}\]
\caption{Defining relations of the braided enveloping algebra $U(\CL)$}
\end{figure}
\proof Formally $U(\CL)$ is the free tensor algebra generated by $\CL$ modulo
these relations with coproduct given by $\Delta$ extended to products as a
braided bialgebra. We have to show that this extension is compatible with the
relations of $U(\CL)$. This is shown in Figure~6. The first equality is the
definition of how $\Delta$ extends to products. The second assumes the
relations in $U(\CL)$. The third is coassociativity and functoriality. The
fourth uses the cocommutativity axiom (L2) applied in reverse. The fifth uses
functoriality and coassociativity again to reorganise. The sixth equality is
(L3). The result the coincides with the extension of $\Delta$ to products when
the relations of $U(\CL)$ are used first. The proof to higher order proceeds
similarly by induction. The proof that $\eps$ also extends to a counit on
$U(\CL)$ is equally straightforward. \endproof
\begin{figure}
\vskip .3in
\epsfbox{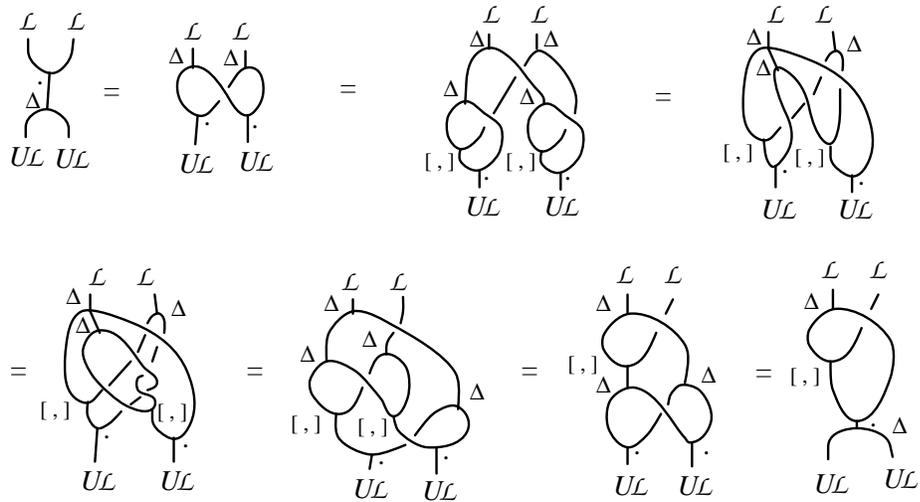}
\caption{Proof that $\Delta$ extends to $U(\CL)$ as a braided bialgebra}
\end{figure}

The motivation here is as follows. In any Hopf algebra one has the identity
$\sum [\xi\o,\eta]\xi\t = \sum \xi\o\o\eta (S\xi\o\t)\xi\t = \xi\eta$. For
example for the usual $U(g)$ with linear coproduct this is
$[\xi,\eta]+\eta\xi=\xi\eta$
as expected. We have a similar definition but without any specific form of
coalgebra, and of course in the braided setting. We conclude with some general
properties of these braided enveloping algebras $U(\CL)$. Following the usual
ideas about Lie algebras representations we have

\begin{defin} A representation of a braided Lie algebra $(\CL,\Delta,\eps,[\ ,\
])$ is an object $V$ and morphism $\alpha:\CL\tens V\to V$ such that the
polarised form of the braided-Jacobi identity (L1) holds. This is shown in
Figure~7 (a). We say that $\CL$ is cocommutative with respect to $V$ if the
polarised form of the cocommutativity axiom
(L2) holds. This is shown in part (b).
\end{defin}
\begin{figure}
\vskip .3in
\epsfbox{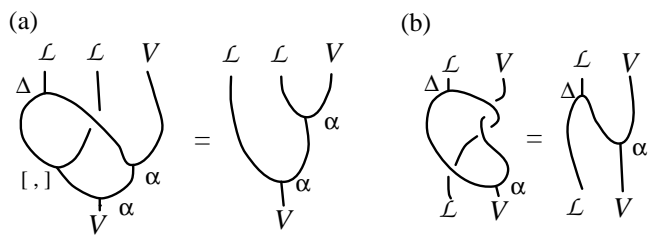}
\caption{Definition (a) of representation of a braided Lie algebra and (b)
cocommutativity with respect to it}
\end{figure}

One can tensor product representations of a braided Lie algebra (using the
coproduct $\Delta$) just as for braided Hopf algebras. The class $\CO(\CL)$ of
representations with respect to which $\CL$ is cocommutative is also closed
under tensor product and braided with braiding given by $\Psi$. The facts are
just as for the representation theory of braided Hopf algebra or
bialgebra\cite{Ma:tra}. The diagrammatic proofs are similar. Alternatively,
these facts follow from the
following proposition that connects representations of $\CL$ to those of
$U(\CL)$ for which the bialgebra theory already developed applies.

\begin{propos} Every representation $(\alpha,V)$ of a braided Lie algebra $\CL$
extends a representation of $U(\CL)$ on $V$. If $\CL$ is cocommutative with
respect to $V$ in the sense of (L2) then $U(\CL)$ is cocommutative with respect
to $V$ in a similar sense (as in \cite{Ma:bra}).
\end{propos}
\begin{figure}
\vskip .3in
\epsfbox{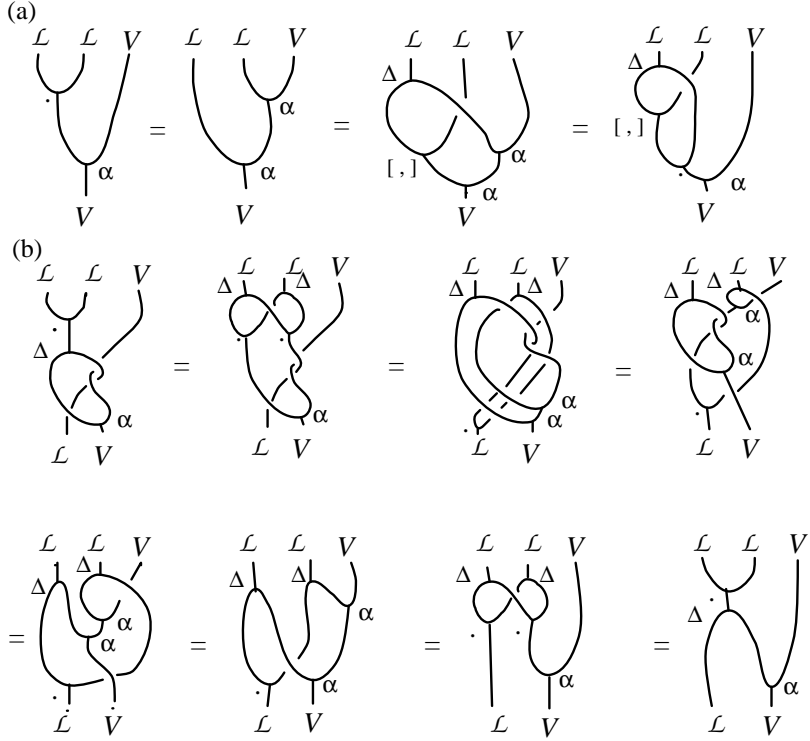}
\caption{Proof that (a) a representation on $V$ extends from $\CL$ to $U(\CL)$
and (b) cocommutativity also extends}
\end{figure}
\proof This is shown in Figure~8. Part (a) verifies that the relations of
$U(\CL)$ are represented correctly. We define the action of $U(\CL)$ by the
repeated application of the Lie algebra action as shown. The representation
axiom in Definition~4.3 ensures that this coincides with the action of $U(\CL)$
if its relations are used first. Part (b) verifies that the resulting action is
cocommutative if the representation is cocommutative. We show it on elements of
$U(\CL)$ with are products of $\CL$. The proof proceeds similarly by induction
to all orders. The first equality uses Proposition~4.2 that $U(\CL)$ is a
bialgebra. The second equality is functoriality to pull one of the products
into the position shown, and that $\alpha$ is a representation for the other
product. The third equality is functoriality again to pull one of the
$\alpha$'s up to the right. We then use the cocommutativity assumption for the
fourth equality, and then again for the fifth. We then use that $\alpha$ is an
action and the bialgebra property of $U(\CL)$ in reverse. \endproof

An important example is of course provided by $[\ ,\ ]$ itself. It was the
model for the definitions and is clearly a representation and $\CL$ is
cocommutative with respect to it. We call it the adjoint representation of
$\CL$ on itself. By the last proposition then, it extends to a representation
(also denoted $[\ ,\ ]$) of $U(\CL)$ on $\CL$ with respect to which $U(\CL)$ is
cocommutative.

\begin{lemma} The adjoint representation $[\ ,\ ]$ of $U(\CL)$ on $\CL$ defined
via Proposition~4.4 obeys an extended form of the braided-Jacobi identity (L1)
and the coalgebra compatibility property (L3) in which the left-most input
$\CL$ in Figure~4 is extended to $U(\CL)$.
\end{lemma}
\begin{figure}
\vskip .3in
\epsfbox{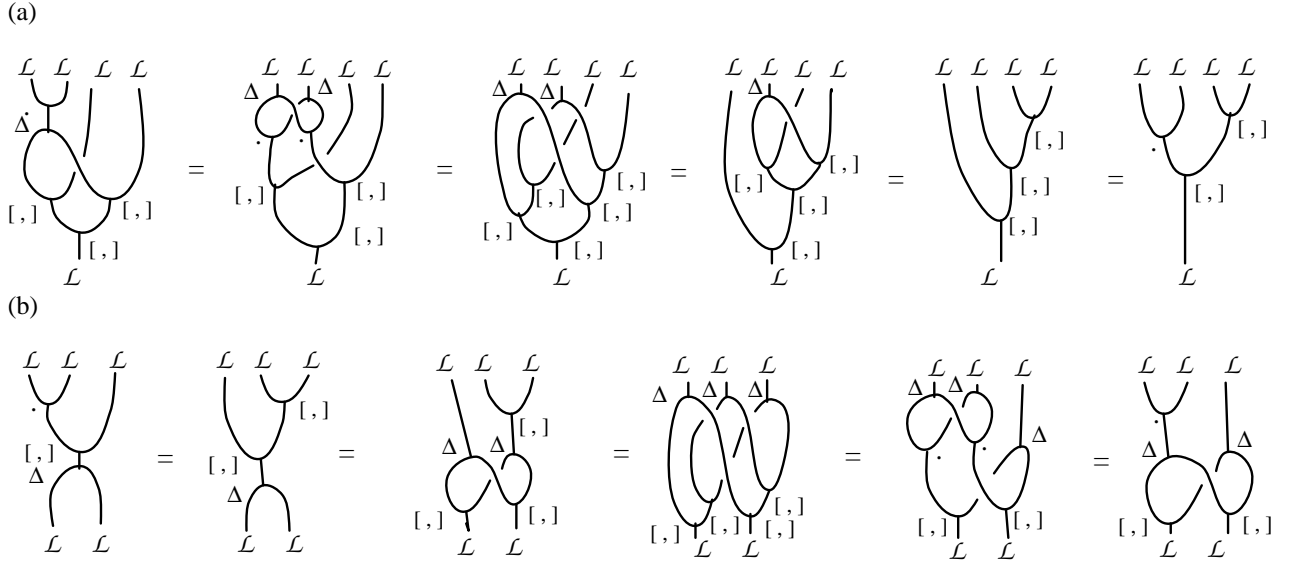}
\caption{Proof that (a) property (L1) and (b) property (L3) extend to
representation $[\ ,\ ]$ of $U(\CL)$ on $\CL$}
\end{figure}
\proof This is shown in Figure~9. Part (a) verifies the extended braided-Jacobi
identity on elements of $U(\CL)$ which are products of $\CL$. The first
equality uses that $U(\CL)$ is a braided bialgebra from Proposition~4.2. The
second that $[\ ,\ ]$ is a representation of $U(\CL)$ as obtained from
Proposition~4.4. We then successively use the braided Jacobi identity axiom
(L1) twice. The final equality uses again that $[\ ,\ ]$ is an action. Exactly
the same proof holds with the elements in $U(\CL)$ is a higher order composite
element, provided only that the result has been proved already at lower orders
so that we can use it for the third and fourth equalities. Hence the result is
proven to all orders by induction. Part (b) is proved in a similar way. We
verify (L3) extended to products in its first input. The first equality is that
$[\ ,\ ]$ is a representation. The second and third successively use (L3). The
fourth then uses that $[\ ,\ ]$ is an action and the fifth that $U(\CL)$ is a
bialgebra. The proof extends to all orders by induction.  \endproof

\begin{propos} The adjoint representation $[\ ,\ ]$ of $U(\CL)$ on $\CL$
defined via Proposition~4.4 extends to a representation on $U(\CL)$ itself as a
braided module algebra. We call it the adjoint action of $U(\CL)$ itself.
$U(\CL)$ remains braided-cocommutative with respect to this action.
\end{propos}

\begin{figure}
\vskip .3in
\epsfbox{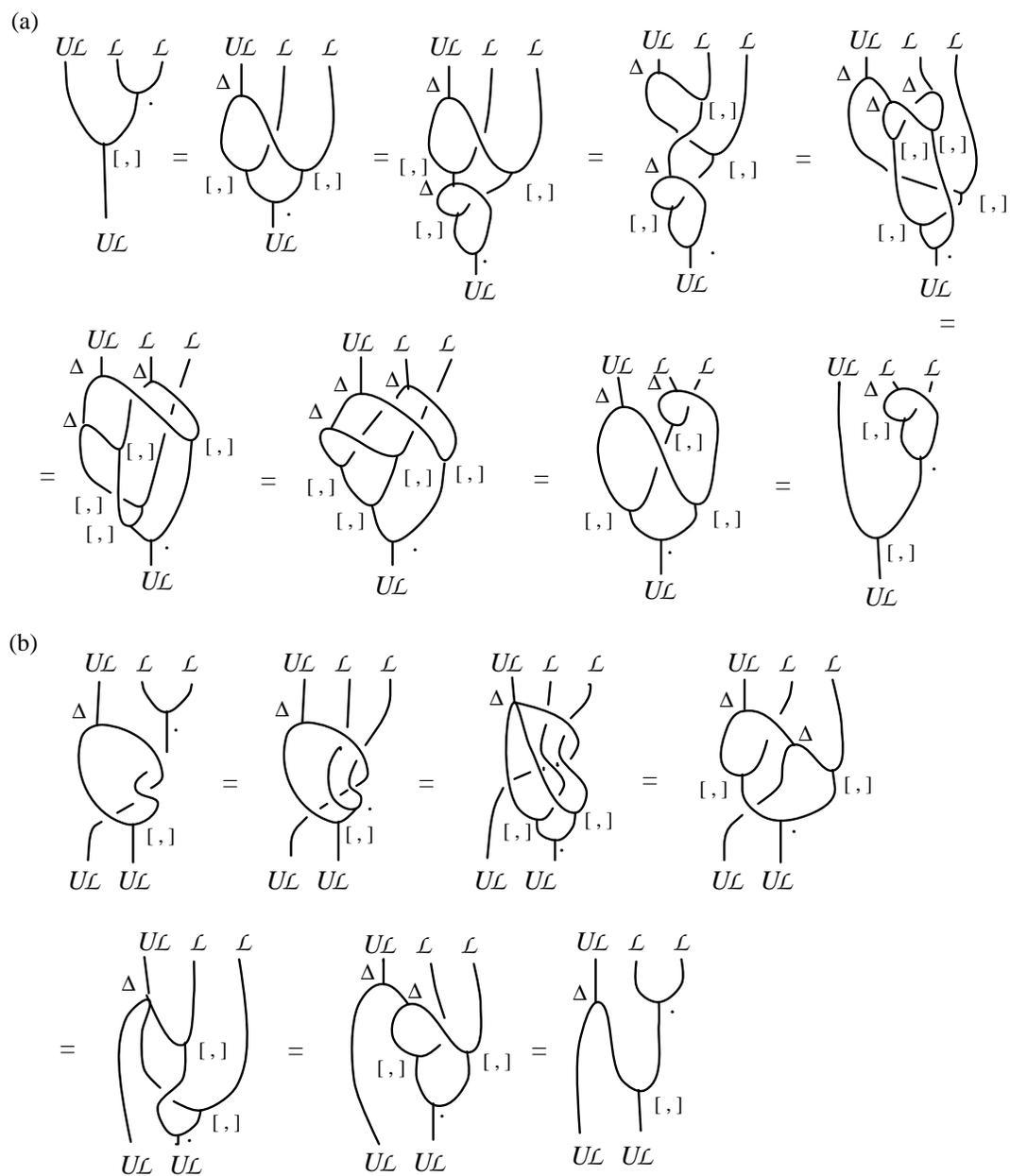}
\caption{Proof that $[\ ,\ ]$ extends to a cocommutative action of $U(\CL)$ on
itself}
\end{figure}
\proof The proof is indicated in Figure~10. We show in part (a) that the
representation constructed in the previous proposition extends consistently as
a braided-module algebra. The first equality is the definition of the extension
in this way. The second uses the relations in $U(\CL)$, the third that $U(\CL)$
acts cocommutatively on $\CL$ from part (b) of the last proposition. The fourth
is axiom (L3). The fifth equality is a reorganization using coassociativity and
functoriality and the sixth is the cocommutativity again. The seventh requires
the preceding lemma that the extended $[\ ,\ ]$ continues to obey a
braided-Jacobi identity as in (L1) but with the first $\CL$ replaced by
$U(\CL)$. Assuming this we see that the result is the same as first using the
relations in $U(\CL)$ and then extending $[\ ,\ ]$ as a braided module algebra.
This proves the result when acting on products or two $\CL$. The proof on
higher products proceeds by induction. Note that in doing this we have to prove
Lemma~4.4 again with the second input of (L1) now also extended to products.
The proof of this is similar to the strategy here (namely consider composites)
and needs the module algebra property of $[\ ,\ ]$ as just proven in Figure~10.
Thus the induction here proceeds hand in hand with this extension of Lemma~4.5.

Part (b) contains the proof that the resulting action of $U(\CL)$ remains
cocommutative on products. The first equality is functoriality while the second
is the module-algebra property just proven. The third and then the fourth each
use the cocommutativity of the $U(\CL)$ action from the preceding proposition.
Coassociativity is expressed by combining branches into multiple nodes (keeping
the order). The fifth equality uses cocommutativity one more time. Finally we
use the module algebra property again to obtain the result. Again the proof on
higher products proceeds in the same way by induction, this time hand in hand
with the extension of the property (L3) in Lemma~4.5 to $U(\CL)$ in its second
input.
This is proven by the same strategy and uses braided-commutativity of the
action of $U(\CL)$ on products of a lower order. \endproof

In the course of the last proof (and using similar techniques) we see that the
braided Jacobi identity  and the coalgebra compatibility property also extend
from $\CL$ to $U(\CL)$. In short, all the properties of $\Ad$ summarized in
Figure~3 hold for this extended $[\ ,\ ]$. We remark that if $\Delta$ on
$U(\CL)$ happens to have an antipode making $U(\CL)$ into a braided Hopf
algebra then
the action $[\ ,\ ]$ indeed coincides with the braided-adjoint action $\Ad$.
This follows easily from the definitions. On the other hand, for a general
coproduct such as the matrix example in the next section, there is no
reason for $U(\CL)$ to be a braided Hopf algebra. It is remarkable that $[\ ,\
]$ nevertheless plays the role of the adjoint action even in this case. Further
properties of these braided enveloping algebras can be developed using similar
techniques to those above.

\begin{figure}
\vskip .3in
\epsfbox{chiLie.eps}
\caption{For coalgebras of the form (a) on $\CX\subset \und 1\oplus\CL$ the
axioms (L1) and (L2) (and also a similar (L3)) of a braided-Lie algebra in
terms of $(\CX,\Delta_1,[\ ,\ ])$ look more familiar. The braided enveloping
algebra in terms of $\CX$ has relations (b)}
\end{figure}

Finally, we note that that $\CL_1=\und 1\oplus\CL\subset U(\CL)$ is also a
coalgebra and closed under the bracket $[\ ,\ ]$ extended as in
Proposition~4.6. Of course the enveloping algebra for this unital coalgebra
$\CL_1$ should be defined without adding another copy of $\und 1$. Otherwise
the construction is just the same as above. Moreover, it may be that another
choice of decomposition of this unital coalgebra $\CL_1$ is possible. For
example $\CL_1=\und 1\oplus\CX$ where $\CX$ is a subobject of the form
\eqn{mordeltachi}{\Delta\chi=\chi\tens
1+1\tens\chi+\Delta_1\chi,\quad\eps\chi=0,\quad\Delta_1:\CX\to\CX\tens\CX}
for $\chi\in \CX$ in concrete cases, and like $\CL$ is closed under $[\ ,\ ]$.
This is expressed in our category by  diagrams as in Figure~11 part (a). In the
other direction if $\Delta_1$ is a morphism which is coassociative (we do not
require it to have a counit) then (\ref{mordeltachi}) defines a coalgebra
structure with $\Delta 1=1\tens 1$ and $\eps 1=1$ in the concrete case. Some
$\CL_1$ of interest below will be of this form and in this case $U(\CL)$ can be
regarded as generated just as well by $\CX$ as $U(\CX)$. From this point of
view a braided Lie algebra of this type is determined by $(\CX,\Delta_1,[\ ,\
])$ in a braided category obeying axioms obtained by putting
(\ref{mordeltachi}) into Figure~4. We use that $[\ ,\ ]$ extends $U(\CL)$ as a
braided-module algebra. The resulting  form of (L1) and (L2) is shown in
Figure~11 and (L3) is obtained in just the same way. In each case nothing is
gained by working in this form (there are just two extra terms) and this is why
we have developed the theory with $(\CL,\Delta,[\ ,\ ])$. On the other hand the
extra terms bring out the sense in which these generators precisely generalise
the usual notion of Lie algebra, with a `braided-correction' $\Delta_1$. Apart
from this we see that (L1) becomes the obvious Jacobi identity in a familiar
form. The enveloping algebra as generated now by the $(\CX,\Delta_1,[\ ,\ ])$
is also of the obvious $\Psi$-commutator form with this $\Delta_1$ correction.

Note that from (L2) in Figure~11 we see that $\Delta_1\ne 0$ if we are to obey
this braided-cocommutativity axiom, unless it happens that $\Psi^2=\id$. Thus,
our notion of braided-Lie algebra in terms of $(\CX,\Delta_1,[\ ,\ ])$ reduces
to precisely the usual notion of Lie-algebra with three terms in the Jacobi
identity etc, only if the category is symmetric and not truly braided. In the
truly braided case there is no advantage to considering the $\CX$ and we may as
well work with the `group-like' generators $\CL$.

\section{Matrix Braided Lie algebras}

The constructions in the last two sections have been rather abstract (and can
be phrased even more formally). In this section we want to show how they look
in a concrete case where the category is generated by a matrix solution of the
QYBE and $\Delta$ has a matrix form.

Firstly, let us recall that our notion of braided Lie algebra is subordinate to
a choice of coalgebra structure on $\CL$. Whatever form we fix determines how
the axioms look in concrete
terms for braided Lie algebras of that type. It need not be the usual implicit
linear form. Thus suppose that $\CL$ is a vector space with basis $\{u^I\}$ say
and fix a coalgebra structure $\und\Delta,\und\eps$ on it. These are determined
in the basis by tensors
\eqn{tenscoalg}{\und\Delta u^I=\Delta^I{}_{JK}u^J\tens u^K,\quad \und\eps
u^I=\delta^I,\quad
\Delta^I{}_{AL}\Delta^A{}_{JK}=\Delta^I{}_{JA}\Delta^A{}_{KL},\quad
\Delta^I{}_{AJ}\delta^A=\delta^I{}_J=\Delta^I{}_{JA}\delta^I}
where $\delta^I{}_J$ is the Kronecker delta function. The underlines on
$\und\Delta$ and $\und\eps$ are to remind is that these are not an ordinary
Hopf algebra coproduct and counit.  Repeated indices are to be summed as usual.
These are obviously the coassociativity and counity axioms in tensor form.

With this chosen coalgebra in the background, the content of Definition~4.1 in
this basis is as follows.

\begin{propos} Let $\CL$ be a vector space with a basis $\{u^I\}$ and coalgebra
$\Delta^I{}_{JK},\delta^I$. Then a braided-Lie algebra on $\CL$ is determined
by tensors ${\bf R}=R^I{}_J{}^K{}_L$ and $c^{IJ}{}_K$ such that $\bf R$ is an
invertible solution of the QYBE and the following three sets of identities hold

(L0a)$\quad \delta^A R^J{}_A{}^I{}_B=\delta^I{}_B\delta^J$ and $\delta^B
R^J{}_A{}^I{}_B=\delta^I\delta^J{}_A$

(L0b)$\quad \Delta^I{}_{MN}\, R^K{}_A{}^N{}_B R^A{}_L{}^M{}_J=
R^K{}_L{}^I{}_A\, \Delta^A{}_{JB}$ and
$\Delta^K{}_{MN}\, R^M{}_A{}^I{}_B\, R^N{}_L{}^B{}_J=R^K{}_B{}^I{}_J\,
\Delta^B{}{}_{AL}$

(L0c)$\quad R^{K}{}_M{}^J{}_B\, R^M{}_L{}^I{}_A\, c^{AB}{}_N=c^{IJ}{}_A\,
R^K{}_L{}^A{}_N$ and
$ R^I{}_A{}^K{}_M\, R^J{}_B{}^M{}_L\, c^{AB}{}_N=c^{IJ}{}_A\, R^A{}_N{}^K{}_L$

(L1)$\quad \Delta^K{}_{PQ}\,  R^I{}_A{}^Q{}_B\, c^{PA}{}_M\, c^{BJ}{}_N\,
c^{MN}{}_L=c^{IJ}{}_A\, c^{KA}{}_L$

(L2)$\quad\Delta^I{}_{PQ}\,  R^{J}{}_A{}^{Q}{}_B\, c^{PA}{}_M\,
R^B{}_L{}^M{}_K=\Delta^I{}_{LB}\,  c^{B J}{}_K$

(L3)$\quad c^{IJ}{}_{A}\, \Delta^{A}{}_{KL}=\Delta^I{}_{MN}\, \Delta^J{}_{PQ}\,
R^{P}{}_A{}^{N}{}_B c^{BQ}{}_L\, c^{MA}{}_K$ and $
c^{IJ}{}_K\delta^K=\delta^I\delta^J$.

\noindent In this case the corresponding braided-Lie algebra structure is
\[ \Psi(u^I\tens u^K)=R^K{}_L{}^I{}_J u^L\tens u^J,\qquad [u^I,u^J]=c^{IJ}{}_K
u^K.\]
The enveloping bialgebra of $\CL$ is generated by the relations
\[ u^Iu^K=\Delta^I{}_{AM}\, R^K{}_B{}^M{}_L\, c^{AB}{}_J\, u^Ju^L.\]
\end{propos}
\begin{figure}
\vskip .3in
\epsfbox{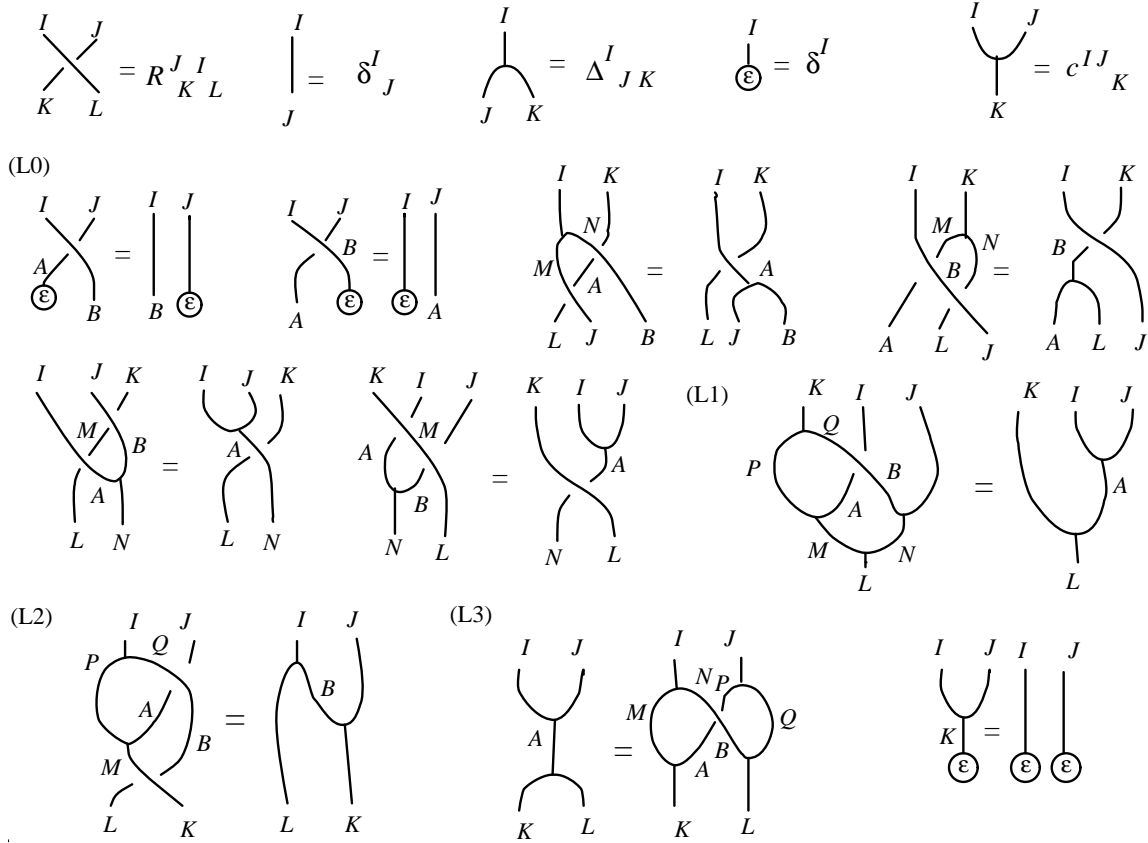}
\caption{Tensor version of braided Lie algebra axioms is obtained by assigning
indices to arcs and tensors as shown}
\end{figure}
\proof We are simply writing the axioms of a braided-Lie algebra as in
Definition~4.1 in our basis. To do this is is convenient to write all
operations as tensors, as we have done already for $\und\Delta$. To read off
the tensor equations simply assign labels to all arcs of the diagram, assign
tensors as shown in Figure~12 and sum over repeated indices. These can be
called braided-Feynman diagrams or braided-Penrose diagrams according to
popular terminology. It is nothing other than our diagrammatic notation in a
basis. The group (L0) are the morphism properties arising from the fact that
$\und\Delta,\und\eps,[\ ,\ ]$ are morphisms in the category and the braiding is
functorial with respect to them, and have been used freely in preceding
sections. In the converse direction, given such matrices, one has to check that
they define a braided Lie algebra. The category in which this lives is the
braided category of left $A$-comodules where (in the present conventions) $A$
is a quotient of the dual-quasitriangular bialgebra $A({\bf R})$. It is in a
certain sense the category generated by ${\bf R}$ and the  braiding is ${\bf
R}$ on the vector space $\CL$ and extended as a braiding to products. The
morphism properties ensure that the relevant maps are morphisms (intertwiners
for the coaction). The other properties needed are (L1)-(L3) which clearly hold
in our basis if the tensor equations hold. Likewise we read-off the relations
for the enveloping bialgebra from Figure~5. \endproof

\note{\eqn{functcoprod}{R^{(i_0,a)}{}_J{}^B{}_L\, R^{(a,i_1)}{}_A{}^K{}_B=
R^K{}_{L}{}^I{}_{(j_0,a_1)}\, \delta^{a_0}{}_{j_1},\quad
R^{(k_0,a)}_B{}^{I}{}_A\, etc -- change R to R_{21} in the following
R^A{}_J{}^{(a,k_1)}{}_L=R^I{}_J{}^K{}_{(b_0,l_1)}\,
\delta^{b_1}{}_{l_0}}
(L1) braided Jacobi identity
\eqn{matjacobi}{ c^{(k_0,a)B}{}_M\, R^{(a,k_1)}{}_A{}^I{}_B\, c^{AJ}{}_N\,
c^{MN}{}_L=c^{IJ}{}_A\, c^{KA}{}_L}
(L2) braided-cocommutativity
\eqn{matcocom}{c^{(i_0,a)B}{}_M\, R^{(a,i_1)}{}_A{}^J{}_B\,
R^M{}_K{}^A{}_L=\delta^{i_0}{}_{l_0}\,  c^{(l_1,i_1) J}{}_K}
(L3) coalgebra compatibility
\eqn{matcompat}{c^{IJ}{}_{(k_0,l_1)}\, \delta^{k_1}{}_{l_0}
=c^{(i_0,a)B}{}_K\, R^{(a,i_1)}{}_A{}^{(j_0,b)}{}_B
c^{A(b,j_1)}{}_L,\qquad c^{IJ}{}_K\delta^K=\delta^I\delta^J}}

To give some concrete examples we now take $\und\Delta$ and $\und\eps$ to be of
matrix form. Thus we work with vector spaces of dimension $n^2$ and let
$\{u^{i_0}{}_{i_1}\}$ denote our basis. Here $I=(i_0,i_1)$ is regarded as a
multiindex. We fix
\eqn{matcoalg}{ \und\Delta u^{i_0}{}_{i_1}=u^{i_0}{}_a\tens u^a{}_{i_1},\quad
\und\eps u^{i_0}{}_{i_1}=\delta^{i_0}{}_{i_1},\quad {\rm i.e.,}\quad
\Delta^I{}_{JK}=\delta^{i_0}{}_{j_0}\delta^{j_1}{}_{k_0}\delta^{k_1}{}_{i_1},\
\delta^I=\delta^{i_0}{}_{i_1}.}
Braided Lie algebras defined with respect to this implicit coalgebra can
naturally be called {\em matrix braided Lie algebras}.

\begin{propos} Let $R\in M_n\tens M_n$ be a bi-invertible solution of the QYBE
(so both $R^{-1}$ and $\widetilde{R}=((R^{t_2})^{-1})^{t_2}$ exist). Then
\[ R^K{}_L{}^I{}_J=R^{i_0}{}_a{}^d{}_{l_0} R^{-1}{}^a{}_{j_0}{}^{l_1}{}_b
R^{j_1}{}_c{}^b{}_{k_1} {\tilde R}^c{}_{i_1}{}^{k_0}{}_d
,\qquad  c^{IJ}{}_K=\widetilde{R}{}^a{}_{i_1}{}^{j_0}{}_b
R^{-1}{}^b{}_{k_0}{}^{i_0}{}_c R^{k_1}{}_{n}{}^c{}_m R^{m}{}_a{}^{n}{}_{j_1}\]
obey the conditions in the preceding proposition and hence define a matrix
braided Lie algebra $(\CL,\Psi,[\ ,\ ])$. Its braided enveloping bialgebra is
the braided-matrices bialgebra introduced in \cite{Ma:exa},
\[ U(\CL)=B(R)\]
with matrix coalgebra $\und\Delta \vecu=\vecu\tens\vecu$, $\und\eps\vecu=\id$.
\end{propos}
\proof In fact, most of the work for this was done in \cite{Ma:exa} where we
proved that $B(R)$ was a braided bialgebra. Apart for an abstract proof (by
transmutation from $A(R)$) we also gave a direct proof in which we verified
directly the relevant identities. This includes most of the above, and the rest
as similar. The matrix $\bf R$ with components $R^K{}_L{}^I{}_J$ was denoted
$\Psi^K{}_L{}^I{}_J$ in \cite{Ma:exa} to avoid confusion with the initial
$R^i{}_j{}^k{}_l$, while the matrix $\CQ$ in \cite{Ma:exa} is basically our
$c^{IJ}{}_K$. The relations of the enveloping algebra are
\[ u^I u^K=c^{(i_0,a)B}{}_J R^K{}_B{}^{(a,i_1)}{}_L\, u^J
u^L=R^{-1}{}^{d}{}_{j_0}{}^{i_0}{}_{a}
R^{j_1}{}_{b}{}^{a}{}_{l_0}R^{l_1}{}_c{}^b{}_{k_1} {\widetilde
R}^c{}_{i_1}{}^{k_0}{}_d\, u^J u^L\]
by multiplying out and canceling some inverses. This is the matrix $\Psi'$ in
\cite{Ma:exa} and defines the relations of $B(R)$. One can move two of the
$R's$ to the left hand side for the more compact form in Section~2. \endproof

Thus the quantum-Lie algebras in Section~2 are successfully axiomatized but
only as braided-Lie algebras. This is therefore the structure that generates
quantum enveloping algebras such as $U_q(g)$. For such standard $R$-matrices
which are deformations of the identity matrix, a more appropriate choice of
generators of $U(\CL)$ is $\chi^I=u^I-\delta^I$. It is standard in the theory
of non-commutative differential calculus to take for the `infinitesimals'
elements such that $\und\eps=0$, and this is what the shift to these generators
achieves. This works fairly generally as follows.

\begin{propos} Let $\CL$ be a braided-Lie algebra in tensor form as in
Proposition~5.1 and $U\CL)$ its braided enveloping algebra with bracket
extended to $U(\CL)$ as in Proposition~4.6. Then the subspace
$\CX=\span\{\chi^I\}\subset U(\CL)$ where $\chi^I=u^I-\delta^I$, is closed
under the braiding and bracket with structure constants
\[ \Psi(\chi^I\tens \chi^K)=R^K{}_L{}^I{}_J\chi^L\tens\chi^J,\qquad
[\chi^I,\chi^J]= (c^{IJ}{}_K-\delta^I\delta^J{}_K)\chi^K\]
and has coalgebra
\[ \und\Delta\chi^I=\chi^I\tens
1+1\tens\chi^I+\Delta^I{}_{JK}\chi^J\tens\chi^K,\quad \und\eps\chi^I=0.\]
\end{propos}
\proof For the braiding we use the morphism properties (L0a) for the counit, to
compute $\Psi(\chi^I\tens\chi^K)$ noting that in its extension to $U(\CL)$ as a
braiding, the braiding of $1$ with anything is trivial (the usual permutation).
For the coproduct $\und\Delta$ we use the counity property in (\ref{tenscoalg})
and that $\und\Delta 1=1\tens 1$ in $U(\CL)$. For the bracket we note that the
extension in Proposition~4.6 is as a braided-module algebra. In particular, $[\
,1]=\und\eps$ and $[1,\ ]=\id$ so that we can compute it on the $\chi^I$.
\endproof

This subspace $\CX$ equally well generates $U(\CL)$ along with $1$, but in
general it is not any more convenient to work than $\CL$ because the coproduct
just has two extra terms and the same term involving $\Delta^I{}_{JK}$. For
example in our matrix setting (\ref{matcoalg}) we have
\[ \und\Delta\chi=\chi\tens 1+1\tens\chi+\chi\tens\chi\]
where the $\chi^I=\chi^{i_0}{}_{i_1}$ are regarded as a matrix. This not better
to work with than our matrix form on $\vecu$. It is however, useful in the
following case.

\begin{corol} Let $(\CL,\Psi,[\ ,\ ])$ be the braided-Lie algebra in
Proposition~5.2 corresponding to a matrix solution $R\in M_n\tens M_n$ of the
QYBE, taken in the form generated by $\CX$ in Proposition~5.3 with its
inherited bracket and braiding. If $R$ is triangular in the sense
$R_{21}R_{12}=1$ then $\Psi$ is a symmetry and the braided-Lie bracket
vanishes,
\[ \Psi^2=\id,\qquad   [\chi^I,\chi^J]=0.\]
Moreover, the enveloping algebra $U(\CL)$ in this case is $\Psi$-commutative in
the sense $\cdot\circ\Psi=\cdot$.

Suppose now that $R$ is not triangular but a deformation $R=R_0+O(\hbar)$ of a
triangular solution $R_0$. If $f^I{}_{JK}$ is the semiclassical part of the
bracket according to
\[ [\chi^I,\chi^J ]=\hbar f^{IJ}{}_K\chi^K+O(\hbar^2)\]
say  on these generators and if we rescale to $\bar\chi^I=\hbar^{-1}\chi^I$
then
\[ [\bar\chi^I,\bar\chi^J]=f^{IJ}{}_K \bar\chi^K + O(\hbar),\quad
\und\Delta\bar\chi^I=\bar\chi^I\tens 1+1\tens\bar\chi^I+O(\hbar)\]
and $f^{IJ}{}_K$ obeys the usual axioms of a $\Psi$-Lie algebra where
$\Psi=\Psi(R_0)$ is the symmetry (this includes usual, super and colour
Lie-algebras etc).
\end{corol}
\proof For the first part we have already pointed out in \cite{Ma:exa} that in
the construction of $B(R)$ the braiding is symmetric if $R$ is triangular and
$c^{IJ}{}_K$ is trivial in the sense $c^{IJ}{}_K=\delta^I\delta^J{}_K$. In any
case these facts follow easily from the explicit forms  of $\Psi,c$ given in
Proposition~5.2. Note that in \cite{Ma:exa} this was interpreted as $B(R)$
being like the $\Psi$-commutative bialgebra of functions on a `space' (like a
super-space), while in the present case we put these observations into
Proposition~5.3 with the interpretation of $B(R)$ as the enveloping algebra of
a $\Psi$-commutative $\Psi$-Lie algebra. For the second part it is clear from
the description of the braided-Jacobi identity and other axioms in Section~4
for the form of the coproduct in Proposition~5.3 that the semiclassical term
$f^{IJ}_K$  obeys precisely the obvious notion of an $R_0$-Lie algebra (where
$R_0$ is triangular, as studied for example in \cite{Gur:yan}\cite{Sch:gen}).
If $R_0=\id$ we have the usual braiding $\Psi$ to lowest order and hence an
ordinary Lie algebra. Another triangular solution is
$(R_S)^i{}_j{}^k{}_l=\delta^i{}_j\delta^k{}_l(-1)^{p(i)p(k)}$ where $p(i)=i-1$
and its deformations in the above framework have super-Lie algebras as their
semiclassical structure.  \endproof

Our formalism is not at all limited to deformations of triangular solutions of
the QYBE, so the matrix braided-Lie algebras in Proposition~5.2 may not
resemble usual Lie algebras or super-Lie algebras or their usual
generalizations. But in the case when $R$ is a deformation of a triangular
solution then they will be deformations of such usual ideas for generalising
Lie algebras when one looks at the generators $\CX$.

We conclude with two of the simplest matrix examples, namely for the initial
$R^i{}_j{}^k{}_l$ given by
\[ R_{gl_2}=\pmatrix{q&0&0&0\cr0& 1& q-q^{-1}& 0\cr0&0&1&0\cr0&0&0&q},\quad
R_{gl_{1|1}}=\pmatrix{q&0&0&0\cr0& 1& q-q^{-1}& 0\cr0&0&1&0\cr0&0&0&-q^{-1}}.\]
Here the rows label $(i,k)$ and the columns $(j,l)$. We denote the matrix
generators as
\[ \vecu=\pmatrix{a&b\cr c&d}\]
and compute from Proposition~5.2. We assume $q^2\ne 1,0$. The corresponding
braidings $\Psi$ and braided enveloping algebras $B(R)$ have already been
computed in \cite{Ma:exa} to which we refer for details of these.

\begin{example}cf\cite{Ma:skl}. Let $R=R_{gl_2}$ be the standard $GL_q(2)$
R-matrix associated to the Jones knot polynomial. A convenient basis for the
corresponding braided-Lie algebra $\CL$ is $\gamma=q^{-2}a+d,\xi=d-a,b,c$
and the non-zero braided Lie-brackets are
\[ [\xi,\xi]=(q^2+1)(q^2-1)^2\xi,\quad [\gamma,\gamma]=(q^{-2}+1)\gamma,\quad
[b,c]=(q^2-1)q^{-2}\xi=-[c,b]\]
\[[\xi,b]=(q^{-2}+1)(q^2-1)b=-q^2[b,\xi],\quad
[\xi,c]=-(q^{-2}+1)(q^2-1)q^{-2}c=-q^{-2}[c,\xi],\quad  \]
\[ [\gamma,\xi]=(q^6+1)q^{-4}\xi,\quad [\gamma,b]=(q^6+1)q^{-4}b,\quad
[\gamma,c]=(q^6+1)q^{-4}c.\]
A convenient basis of $\CX$ is $\xi,b,c$ and $\gamma-\und\eps(\gamma)$ which we
rescale by a uniform factor $(q^2-1)^{-1}$ to a basis $\bar\xi,\bar b,\bar
c,\bar\gamma$. Then
the braided-Lie algebra takes the form
\[[\bar \xi,\bar b]=(q^{-2}+1)\bar b=-q^2[\bar b,\bar \xi],\quad [\bar \xi,\bar
c]=-(q^{-2}+1)q^{-2}\bar c=-q^{-2}[\bar c,\bar \xi],\quad [\bar b,\bar
c]=q^{-2}\bar\xi=-[\bar c,\bar b] \]
\[ [\bar\xi,\bar\xi]=(q^{4}-1)\bar\xi,\quad
[\bar\gamma,\bar\xi]=(1-q^{-4})\xi,\quad [\bar \gamma,\bar b]=(1-q^{-4})\bar
b,\quad [\bar\gamma,\bar c]=(1-q^{-4})\bar c.\]
with zero for the remaining six brackets. As $q\to 1$ the braiding $\Psi$
becomes the usual transposition and the  space $\CX$ with its bracket becomes
the Lie algebra $sl_2\oplus u(1)$. The bosonic generator $\bar\gamma$ of the
$U(1)$ decouples completely in this limit.
\end{example}
\proof This is from the definition on Proposition~5.2. It is similar to the
computation of the action of $U_q(sl_2)$ on for the degenerate Sklyanin algebra
in \cite{Ma:skl}. We computed $B(R)$ in \cite{Ma:eul}\cite{Ma:exa} and already
noted the importance of the element $d-a=\xi$, and that the element
$q^{-2}a+d=\gamma$ as bosonic and central in $B(R)$. It is remarkable that its
braided Lie bracket is not entirely zero even though the action of $U_q(sl_2)$
on it is trivial. The shift to the barred variables follows the general theory
explained above since $R$ here is a deformation of a triangular solution
(namely the identity). To compute the brackets $[\bar\gamma,\ ]$ we note that
$\und\eps(\gamma)=(q^{-2}+1)$ and that the bracket obeys $[1,\ ]=\id$ and $[\
,1]=\und\eps$. Hence
\[[\gamma-\und\eps(\gamma),b]=[\gamma,b]
-(q^{-2}+1)[1,b]=((q^6+1)q^{-4}-(q^{-2}+1))b=(q^2-1)(1-q^{-4})b\]
\[[\gamma-\und\eps(\gamma),\gamma-\und\eps(\gamma)]
=[\gamma,\gamma]-(q^{-2}+1)[1,\gamma]=0\]
etc. The other computations are similar. The braiding $\Psi$ and the structure
of the enveloping algebra are in \cite{Ma:exa}
\endproof

Note that braided enveloping bialgebra $U(\CL)$ in terms of these rescaled
generators must in the limit $q\to 1$ tend to $U(gl_2)$. It can be called
$BU_q(gl_2)$ because it is a braided object. We have identified it in
\cite{Ma:skl} as the degenerate Sklyanin algebra. On the other hand this same
$B(R)$ in terms of the original generators $\vecu$ tends to the commutative
algebra  algebra generated by the co-ordinate functions on the space of
matrices $M_2$ which was our original point of view in
\cite{Ma:eul}\cite{Ma:exa}. Thus for generic $q$ we can think of the braided
bialgebra $U(\CL)=B(R)$  from either of these points of view. The same applies
in the next example where we took the view in \cite{Ma:exa}
that $B(R)$ tends as $q\to 1$ to the super-bialgebra of super-matrices
$M_{1|1}$. This time, after rescaling it becomes in the limit the
super-enveloping algebra $U(gl_{1|1})$.

\begin{example} Let $R=R_{gl_{1|1}}$ be non-standard R-matrix associated to the
Alexander-Conway
knot polynomial. A  convenient basis for the corresponding braided-Lie algebra
$\CL$ is $a,\xi=d-a,b,c$
and the non-zero braided-Lie brackets are
\[ [b,c]=-(q^2-1)\xi=q^{2}[c,b],\quad [b,a]=(q^2-1)q^{-2}b,\quad
[c,a]=-(q^2-1)c\]
\[ [\xi,a]=-(q^2-1)^2q^{-2}\xi,\quad [a,\xi]=\xi,\quad [a,a]=a,\quad
[a,b]=q^{-2}b,\quad [a,c]=q^2c\]
A convenient basis for $\CX$ is $a-1,b,c,\xi$ which we rescale by a uniform
factor $(q^2-1)^{-1}$ to obtain a basis $\bar a,\bar b,\bar c,\bar\xi$. Then
the braided-Lie algebra takes the form
\[ [\bar b,\bar c]=-\bar \xi=q^{2}[\bar c,\bar b],\quad [\bar \xi,\bar
a]=(q^{-2}-1)\bar \xi,\quad [\bar a,\bar b]=-q^{-2}\bar b=-[\bar b,\bar
a],\quad [\bar a,\bar c]=\bar c=-[\bar c,\bar a]\]
with zero for the remaining nine brackets. As $q\to 1$ the braiding $\Psi$ is
such that $\CX$ becomes a super-vector space with $\bar a,\bar\xi$ even degree
(bosonic) and $\bar b,\bar c$ odd degree (fermionic), and its bracket becomes
that for the super-Lie algebra $gl_{1|1}$.
\end{example}
\proof This is by direct computation from Proposition~5.2. The enveloping
algebra $B(R)$ was studied in \cite{Ma:exa} where we identified the element
$\xi=d-a$ as bosonic and central. The passage to the barred variables follows
the same steps as the previous example. The braiding $\Psi$ and the structure
of the enveloping algebra are in \cite{Ma:exa}. \endproof

This example tends as $q\to 1$ to a super-Lie algebra, as it must from the
general theory described above. This is because $R$ tends to the matrix $R_S$
which is the critical limit point for super-Lie algebras. The corresponding
braiding $\Psi$ for this is the usual super-transpositions. It is a triangular
solution of the QYBE and all its deformations lead by the above to super-Lie
algebras.

In this way we see that our general R-matrix construction for braided algebras
unifies the notions of Lie algebras and super-Lie algebras, colour-Lie algebras
etc., into a single framework.
These usual notions are the semiclassical part of the structure as we approach
a certain subset (the triangular solutions) in the moduli space of all
solutions of the quantum Yang-Baxter equations. On the other hand we are not at
all tied in principle to such
usual deformations. For example if we consider our braided-Lie algebras at
other points $R$ in the moduli space it is natural to call the corresponding
semi-classical structures R-Lie algebras. They control the deformations of
$B(R)$ (the enveloping algebra at $R$).  One possible application may be that
by solving some kind of R-classical Yang-Baxter equation for general $R$ (based
on an R-Lie algebra) one should be able to exponentiate to paths in the moduli
space. Moreover, the usual quantum groups are precisely quotients of such
enveloping algebras so we have the possibility of connecting them by paths in
the moduli space. This is a problem for further work.

\section{Braided-Vector Fields}

In this section we show that the braided enveloping algebras $U(\CL)$ act quite
naturally as braided-vector fields on braided-function algebras. We have
already seen one example namely the bracket $[\ ,\ ]$ consisting of one copy of
$U(\CL)$ acting on another. In the construction of Proposition~5.2 the braided
enveloping algebra can also be thought of as the braided-matrix function
algebra and we do so for the copy of $U(\CL)$ which is acted upon. The
vector-fields in this case are (in a braided-group quotient) those induced by
the adjoint action. In this section we give by contrast vector fields
corresponding to the right action on functions induced by left-multiplication
in the group (the right regular representation).

In the case of usual matrix groups recall that these vector fields are
literally given by matrix multiplication of the Lie algebra elements realised
as matrices on the group elements. Thus, if $u^i{}_j$ are the matrix
co-ordinate functions on the matrix group in the defining representation
$\rho$, $g$ a group element and $\xi$ a Lie algebra element, we have
\[ (u^i{}_j\ra\xi)(g)=u^i{}_j(\xi g)=\rho(\xi)^i{}_k u^k{}_j(g).\]
Our constructions in this section give in the matrix case of Proposition~5.2
precisely a $q$-deformation of this situation. We realise our matrix
braided-Lie algebras concretely as matrices acting by a deformation of matrix
multiplication. This is in marked contrast to usual quantum groups, but mirrors
well the situation for super groups and super-matrices and their super-Lie
algebras.

Our strategy to obtain this result is to go back to the abstract situation
where we have a braided Hopf algebra $B$ in a braided category, formulate the
construction there and afterwards compute its matrix form. Because the relevant
braided matrices and braided groups that concern us are related (in the nice
cases) to quantum groups by a process of transmutation, we obtain on the way
vector-fields on quantum groups also.

The general construction of the regular representation proceeds in our
categorical setting in Section~3 along the same lines as the braided-adjoint
action. Namely, one writes the usual group or Hopf algebra construction in
diagrammatic form. Note that the coproduct of $B$ encodes the group
multiplication law if $B$ is like the algebra of functions on a group. The
evaluation of this against an element of the dual $B^*$ is then like the action
of the enveloping or group algebra in the regular representation. This gives
the following construction.

\begin{propos} Let $B$ be a Hopf algebra in a braided category as in Section~3
and suppose that it has a dual $B^*$. Then $B^*$ acts on $B$ from the right as
depicted in the box in Figure~13. Moreover, the action  respects the product on
$B$ in the sense that $B$-becomes a $B^*$-module algebra. We call this the
right-regular action.
\end{propos}
\begin{figure}
\vskip .3in
\epsfbox{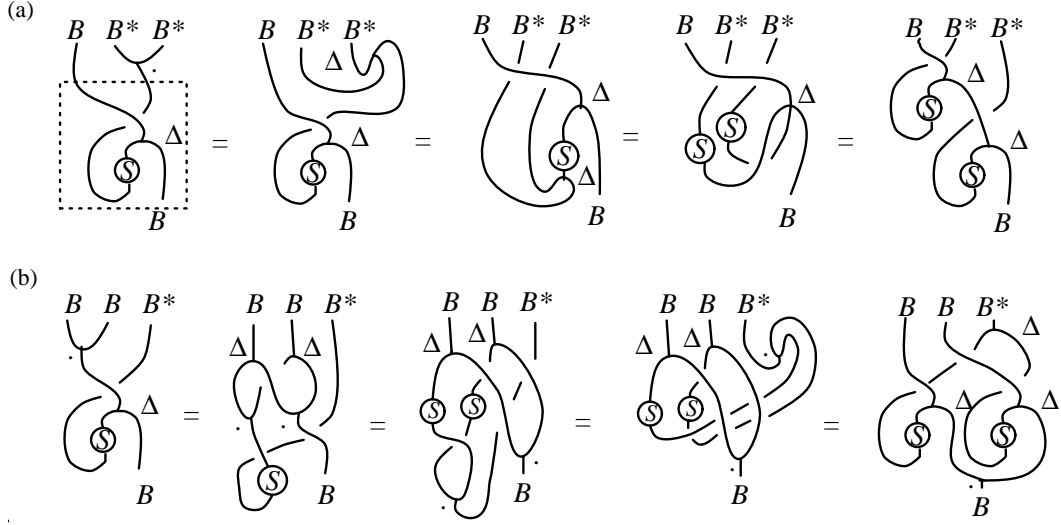}
\caption{The braided right action of $B^*$ on $B$ is shown in the box. It is
(a) a right action and (b) a right braided module algebra}
\end{figure}
\proof  Here $B^*$ assumes that our category is equipped with dual objects (in
this case left duals) and the cup and cap denote evaluation $\ev: B^*\tens
B\to\und 1$ and coevaluation $\coev:\und 1\to B\tens B^*$ respectively. They
obey a natural compatibility
\eqn{ev-coev}{(\ev\tens\id)(\id\tens\coev)=\id,\qquad
(\id\tens\ev)(\coev\tens\id)=\id} which in diagrammatic form says that certain
horizontal double-bends can be pulled straight. The unusual ingredient in the
right action is the braided antipode $S$ which converts a left action to a
right action and is needed in the strictly braided case for the module algebra
property to work out without getting tangled. The proof that this is an action
is in part (a). The first equality is the definition of the product in $B^*$ in
terms of the coproduct in $B$. In terms of maps this is equivalent to the
characterization
\eqn{B*}{\ev\circ(\id\tens\ev\tens\id)\circ(\id\tens
\Delta_B)=\ev\circ(\cdot_{B^*}\tens\id).}
The second equality is the double-bend cancellation property of left duals. We
then use that the fact that the braided-antipode is a braided anti-coalgebra
map and functoriality to recognize the result. That this makes $B$ a braided
right module algebra is shown in part (b). The first equality is the bialgebra
axiom, the second is the fact that the braided-antipode is a
braided-antialgebra map, the third functoriality and the fourth the definition
of the coproduct in $B^*$ in terms of the product in $B$. This is determined in
a similar way to (\ref{B*}) via pairing by $\ev$. An introduction to the
methods is in \cite{Ma:introp}. \endproof

Now let $H$ be an ordinary quasitriangular Hopf algebra dually paired as as in
Section~2 with suitable dual $A$. There are associated braided groups $B(H,H)$
and $B(A,A)$ corresponding to these by transmutation\cite{Ma:bra}\cite{Ma:eul}.
They can both be viewed in the braided category of $H$-modules and as such
$B(H,H)=B(A,A)^*$ at least in the finite dimensional case. We can therefore
apply the above diagrammatic construction and compute the action of $B(H,H)$ on
$B(A,A)$. The resulting formulae can also be used with care even in the
infinite dimensional case.

\begin{propos} The canonical right-action of $B(H,H)$ on $B(A,A)$ (the braided
group of function algebra type) comes out as
\[a\ra b=\sum <\CR\ut\la b, a\o>a\t<\CR\uo,a\th>,\quad a\in B(A,A),\ b\in
B(H,H).\]
This makes $B(A,A)$ into a right braided $B(H,H)$-module algebra in the
category of left $H$-modules.
\end{propos}
\proof We compute from the formulae for $B(H,H)$ in \cite{Ma:bra} using
standard Hopf algebra techniques. Its product is that same as that of $H$ and
it lives in the stated category by the quantum adjoint action $\la$. We need
the explicit formulae
\[ \und\Delta b=\sum b\o (S\CR\ut)\tens \CR\uo\la b\t,\quad \und S b=
u^{-1}(SR\ut)(Sb)\CR\uo\]
for the braided coproduct and braided-comultiplication, where $u=\sum
(S\CR\ut)\CR\uo$ implements the square of the antipode. Finally, $B(A,A)$ has
the same coproduct as $A$, transforms under the quantum coadjoint action and is
dually paired by the map $B(H,H)\to B(A,A)^*$ given by $b\mapsto <Sb,\ >$.
Armed with these explicit formulae we compute  the box in Figure~13 as
\align{a\ra b &&\nqquad =\sum <S(\CR\ut\la \und S b),\CR\uo\o\la
a\o>\CR\uo\t\la a\t\\
&&=\sum <(S\CR\uo\o)(S(\CR\ut\la\und S
b))\CR\uo\t,a\o>a\th<S\CR\uo\th,a\t><\CR\uo{}_{(4)},a{}_{(4)}>\\
&&=\sum <(S\CR\uo)S(\CR\ut\CR'\ut\la \und S b),a\o> a\t <\CR'\uo,a\th>\\
&&=\sum <(S\CR\uo)S(\CR\ut\la \und S(\CR'\ut\la b)),a\o> a\t <\CR'\uo,a\th>\\
&&=\sum <\CR'\ut\la b,a\o> a\t <\CR'\uo,a\th>.}
Here the first equality follows from the form of $\und\Delta$ and of the
braiding $\Psi$ in the category of $H$-modules (it is given by the action of
$\CR$ followed by usual permutation). The second equality puts the coadjoint
action as an adjoint action on the other side of the pairing in one case, and
computes it in the other case. The third equality writes the coproduct in $A$
as a product in $H$ and cancels using the antipode axioms. We also used the
axioms of a quasitriangular structure (\ref{quasitr}). The fourth uses that
$\und S$ is a morphism in the category (an intertwiner). Finally we use for the
last equality the definition of $\und S$ in the reverse form
\[ \sum (\CR\ut\la \und S b)\CR\uo=u^{-1}(Sb)u=S^{-1}b,\qquad\forall b\in
B(H,H)\]
easily obtained from the formula above. We apply this to the element
$\CR'\ut\la b$.

{}From the general categorical construction above, we know that this right
action has all the properties of a braided-module algebra. One can (in
principle) verify some of these directly. For example, that $\ra$ as stated is
a morphism in the category means
\eqn{racov}{ h\la(a\ra b)=\sum (h\o\la a)\ra(h\t\la b),\qquad \forall h\in H}
which can be verified directly using the standard properties of quasitriangular
Hopf algebras as can that $\ra$ is indeed an action. The module algebra
property is more difficult to see directly. \endproof

In the infinite-dimensional case we take here the category of $A$-comodules and
write $\CR$ as a dual-quasitriangular structure $A\tens A\to \C$. For $H$ we
can then take for example $U_q(g)$ in FRT form. The braided-version $B(H,H)$
has isomorphic algebra and coincides in this factorizable case to a quotient of
$U(\CL)$ for the corresponding braided Lie algebra $\CL$. For $A$ we can take
the quantum function algebra $\CO_q(G)$ and as seen in
\cite{Ma:eul}\cite{Ma:exa} its corresponding braided version $B(A,A)$ is a
quotient of $B(R)$. In this case we can compute the action in Proposition~6.2
as
\align{u^i{}_j\ra l^k{}_l&=& < (S\CR\ut)\la
l^k{}_l,t^i{}_a>u^a{}_b<S\CR\uo,t^b{}_j>\\
&&=<Sl^-{}^b{}_j\la l^k{}_l,t^i{}_a>u^a{}_b=<\widetilde{R}{}^a{}_j{}^k{}_m
l^m{}_n R^b{}_a{}^n{}_l,t^i{}_a>u^a{}_b\\
&&=u^a{}_b \widetilde{R}{}^m{}_j{}^k{}_n Q^n{}_p{}^i{}_a R^b{}_m{}^p{}_l}
using the notations in Section~2. We used (\ref{Slm-Ad-l}) and the definition
of $l^-$ in terms of the quasitriangular structure $\CR$. Moreover, we know
that the construction the covariant under a background copy of $U_q(g)$ in the
sense of (\ref{racov}) with action as in (\ref{lpm-Ad-l} on $\vecu$. Clearly
the same constructions apply for any $R$ which is sufficiently nice that we
have a factorizable quantum group in the picture. On the other hand, we are now
ready to verify directly that this whole construction lifts to the bialgebra
level. It is quite natural at the level of braided-Lie algebras.

\begin{propos} Let $R$ be a bi-invertible solution of the QYBE as in
Proposition~5.2 and $\CL$ the braided Lie algebra introduced there. Let $B(R)$
be the braided-matrix bialgebra. Then $\CL$ acts from the right on the algebra
of $B(R)$ by braided-automorphisms ($B(R)$ is a right-braided module algebra
for the action of $(\CL,\Delta)$). We write $\ra
u^{i_0}{}_{i_1}=\overleftarrow{\del}{}^{i_0}{}_{i_1}=\overleftarrow{\del}{}^I$
for the corresponding operators. Then
\[u^{i_0}{}_{i_1}\overleftarrow{\del}{}^{j_0}{}_{j_1}=  u^{k_0}{}_{k_1}
\widetilde{R}{}^c{}_{i_1}{}^{j_0}{}_b Q^b{}_a{}^{i_0}{}_{k_0}
R^{k_1}{}_c{}^a{}_{j_1}\]
and the extension is according to the braided-Leibniz rule
\[ (ab)\overleftarrow{\del}{}^{i_0}{}_{i_1}= a\cdot\Psi(b\tens
\overleftarrow{\del}{}^{i_0}{}_k)\overleftarrow{\del}{}^k{}_{i_1},\qquad\forall
a,b\in B(R).\]
\end{propos}
\proof We no longer need a quantum group, but if there is one it remains a
background covariance of the system as above. For our direct verification it is
convenient to write the action compactly as
\eqn{right-vec-mat}{ \vecu_1 R_{12}\overleftarrow{\del_2}=Q_{21}\vecu_1
R_{12}=\rho_1(\vecu_2)\vecu_1R_{12}}
where $\rho$ is the fundamental representation of $U(\CL)$ defined in
Lemma~2.5. From this it is clear that the operators $\overleftarrow{\del}$ are
truly a representation of $U(\CL)$ as required, and hence also of $\CL$ in the
sense of Definition~4.3. Next we need to check that the extension of this
action to products as a right-braided module algebra,
\eqn{mat-leib-B(R)}{(\vecu_1 R_{23}^{-1}\vecu_2R_{23})
\overleftarrow{\del}{}_3=(\vecu_1\overleftarrow{\del}{}_3) (R_{23}^{-1}\vecu_2
R_{23}\overleftarrow{\del}{}_3)}
etc, respects the relations of $B(R)$. In proving this it is convenient to
insert some
$R$-matrices and prove compatibility with the relations in an equivalent form.
Thus,
\align{&&\nqquad (R_{21}R_{13}^{-1}\vecu_1 R_{13}R_{12}R_{23}^{-1}\vecu_2
R_{23})\overleftarrow{\del_3}=(R_{23}^{-1}R_{13}^{-1} R_{21}\vecu_1
R_{12}\vecu_2 R_{13}R_{23})\overleftarrow{\del_3}\\
&&=(R_{23}^{-1}R_{13}^{-1} \vecu_2R_{21}\vecu_1
R_{12}R_{13}R_{23})\overleftarrow{\del_3}=(R_{23}^{-1}\vecu_2
R_{23}R_{21}R_{13}^{-1}\vecu_1R_{13})\overleftarrow{\del_3}R_{12}\\
&&=(R_{23}^{-1}\vecu_2 R_{23}R_{21}\overleftarrow{\del_3})(R_{13}^{-1}\vecu_1
R_{13}\overleftarrow{\del_3})R_{12}=R_{32}R_{31}\vecu_2 R_{21}\vecu_1
R_{23}R_{13}R_{12}.}
Here the first equality is a few applications of the QYBE, the second the
relations in $B(R)$ and the third the QYBE again (this combination is the
relations of $B(R)$ transformed under $l^+\la$ as in Section~2). The fourth
equality is our supposed extension according to (\ref{mat-leib-B(R)}). We
compute the derivatives from (\ref{right-vec-mat}) and use the QYBE for the
fifth. On the other hand if we begin from the same starting point and use
(\ref{right-vec-mat}) we have
\[(R_{21}R_{13}^{-1}\vecu_1
R_{13}R_{12}\overleftarrow{\del_3})(R_{23}^{-1}\vecu_2
R_{23}\overleftarrow{\del_3})=R_{32}R_{31}R_{21}\vecu_1 R_{12}\vecu_2
R_{13}R_{23}\]
which gives the same result as above using the relations in $B(R)$. From this
it follows that these relations are compatible with the action of $\CL$. The
direct computation with tensor indices (rather than the compact notation) is
also possible. \endproof

This is the natural right action of $B(R)$ regarded as a braided enveloping
algebra $U(\CL)$ on itself regarded as a braided function algebra. Just as in
Corollary~5.4, it is trivial if $R$ is triangular. It is natural in this case
to define the action of the infinitesimal generators $\chi^I$. This is
$\ra\chi^I=\overleftarrow{\del^I}-\overleftarrow{\delta^I}=\overleftarrow{D^I}$
say, and from (\ref{right-vec-mat}) it is clear that it vanishes if $R$ is
triangular.

\begin{corol} If $R$ is a solution of the QYBE such that $R=R_0+O(\hbar)$ where
$R_0$ is a triangular solution, then $\overleftarrow{D^I}=O(\hbar)$ and the
action of the rescaled generators
$\ra\bar\chi^I=\hbar^{-1}\overleftarrow{D^I}=\overleftarrow{\bar D^I}$ is a
usual $\Psi$-derivation. Here $\Psi$ is from Proposition~5.2 with $R=R_0$ and
is a symmetry.
\end{corol}
\proof As in Figure~11, we compute the form of the right-module algebra
property in Figure~13 for the form of $\Delta$ on the $\bar\chi$. Explicitly,
\eqn{chi-leib}{(ab)\overleftarrow{\bar D^I}=a(b\overleftarrow{\bar
D^I})+a\Psi(b\tens \overleftarrow{\bar D^I}) + \hbar a\Psi(b\tens
\overleftarrow{\bar D}{}^{(i_0,k)})\overleftarrow{\bar D}{}^{(k,i_1)}.}
The last $\Delta_1$ term enters at order $\hbar$ as does the deformation of the
braiding. Hence to lowest order the $\hbar^{-1}\overleftarrow{D^I}$ obey the
usual axioms of a right-vector field in a symmetric monoidal category.
 \endproof

Recall that it is these rescaled generators that behave like usual Lie algebras
or super-Lie algebras etc to lowest order as we approach the critical variety
of triangular solutions of the QYBE. We see that in this case it is exactly
these that act on the braided matrices $B(R)$ in this corollary. Here $B(R)$
itself becomes in the triangular limit the $\Psi$-commutative algebra of
functions on some kind of matrix space. Moreover, these constructions work at
the braided-group level so the underlying space here can be regarded as some
kind of group-manifold in the sense of a supergroup or ordinary group etc.

\begin{example} For $R_{gl_2}$ as in Example~5.5 the matrix-braided vector
fields are
\[\scriptstyle\overleftarrow{\del}{}^1{}_1=
\pmatrix{q^2&0&0&(q-q^{-1})^2\cr0&q^2&0&0\cr0&0&1&0\cr 0&0&0&1},\qquad
\overleftarrow{\del}{}^1{}_2=\pmatrix{0&0&1-q^{-2}&0\cr0&0&0&q^2-1\cr0&0&0&0\cr
0&0&0&0}\]
\[\scriptstyle\overleftarrow{\del}{}^2{}_1=
\pmatrix{0&-(1-q^{-2})^2&0&0\cr0&0&0&0\cr q^2-1&0&0& (q-q^{-1})^2\cr
0&1-q^{-2}&0&0},\quad
\overleftarrow{\del}{}^2{}_2=
\pmatrix{q^2+q^{-2}-1&0&0&-(1-q^{-2})^2\cr0&q^2+q^{-2}-1&0&0
\cr0&0&q^2&0\cr 0&0&0&q^2}.\]
{}From this we obtain the action of the rescaled generators $\bar\chi$ as
\[ \pmatrix{a&b\cr c& d}\ra{\bar \xi} =\pmatrix{-q^{-2}a&-q^{-2}b\cr
c&d+(q^{-4}-1)a},\quad
\pmatrix{a&b\cr c& d}\ra{\bar b}=\pmatrix{0&0\cr q^{-2}a&b}\]
\[ \pmatrix{a&b\cr c& d}\ra{\bar c}=\pmatrix{c&q^{-2}d-(1-q^{-2})q^{-2}a\cr
0&(1-q^{-2})c},\quad   \pmatrix{a&b\cr c& d}\ra{\bar \gamma}=\pmatrix{a&b\cr
c&d}.\]
As $q\to 1$ this becomes the usual right action of the lie algebra $gl_2$ on
the co-ordinate functions of $M_2$.
\end{example}
\proof This is by direct computation from Proposition~6.3. The
$\overleftarrow{\del}$ act on the row vector $(a,b,c,d)$ by the matrices shown.
{}From this by subtracting the identity matrix from
$\overleftarrow{\del}{}^1{}_1$ and $\overleftarrow{\del}{}^2{}_2$ we obtain the
action of the $\chi^i{}_j$ variables. This then gives the action of the
rescaled basis $\bar\xi,\bar b,\bar c,\bar\gamma$, where the rescaling is by
$(q^2-1)^{-1}$ as before. These also act by $4\times 4$ matrices on the
generators of $B(R)$, which we write now in a more explicit form as shown. From
this explicit form we see that as $q\to 1$ the actions become
\[\pmatrix{a&b\cr c&d}\ra\bar\xi=\pmatrix{-1&0\cr0&1}\pmatrix{a&b\cr c&d},\quad
\pmatrix{a&b\cr c&d}\ra\bar b=\pmatrix{0&0\cr1&0}\pmatrix{a&b\cr c&d},\quad
\pmatrix{a&b\cr c&d}\ra\bar c=\pmatrix{0&1\cr0&0}\pmatrix{a&b\cr c&d}\]
which is the usual action of the $sl_2$ generators by left-invariant vector
fields on the functions algebra of $SL_2$ or $M_2$ as here.

Note that at the level of $U(\CL)$ and its action on $B(R)$, the choice of
normalization of this initial $R$ is not important. It does not change the
algebras and simply scales the $\overleftarrow{\del}$ in Proposition~6.3. On
the other hand since the action of $1$ is not scaled, the action of the $\chi$
generators can change more significantly. For the present example the so-called
quantum-group normalization for the present $R$-matrix requires an additional
factor $q^{-\h}$ in $R_{gl_2}$. This means a uniform factor $q^{-1}$ in the
$\overleftarrow{\del}$ as well as for the $\ra\bar b,\ra\bar c,\ra\bar\xi$,
while $\ra\bar\gamma$ now acts by a different multiple of the identity. This
normalization is the one needed for the representation of $BU_q(gl_2)$ to
descend to the quantum group $U_q(sl_2)$, for which $\gamma$ becomes
proportional to its quadratic Casimir. On the other hand, we are not tied to
this consideration and have retained the normalization that seems more suitable
for the braided enveloping bialgebra. \endproof

We see that when $q\to 1$ the action of the braided-vector fields becomes the
usual action by left-multiplication of the Lie algebra on the co-ordinate
functions, as it must by the constructions above. On the other hand for general
$q$ or other non-standard $R$-matrices it is not possible to write the actions
of our braided-vector fields as a matrix product of the Lie algebra matrix on
the group matrix. This problem is well-known even in the case of super-Lie
algebras acting by super-vector-fields.

\begin{example} For $R_{gl_{1|1}}$ as in Example~5.6 the matrix-braided vector
fields are
\[\scriptstyle\overleftarrow{\del}{}^1{}_1=
\pmatrix{q^2&0&0&(q-q^{-1})^2\cr0&q^2&0&0\cr0&0&1&0\cr 0&0&0&1},\qquad
\overleftarrow{\del}{}^1{}_2=\pmatrix{0&0&1-q^{-2}&0
\cr0&0&0&q^{-2}-1\cr0&0&0&0\cr 0&0&0&0}\]
\[\scriptstyle\overleftarrow{\del}{}^2{}_1
=\pmatrix{0&(q-q^{-1})^2&0&0\cr0&0&0&0\cr q^2-1&0&0& (q-q^{-1})^2
\cr 0&1-q^2&0&0},\quad
\overleftarrow{\del}{}^2{}_2=\pmatrix{q^2+q^{-2}-1&0&0&(q-q^{-1})^2
\cr0&q^2+q^{-2}-1&0&0\cr0&0&q^{-2}&0\cr 0&0&0&q^{-2}}.\]
{}From this we obtain the action of the rescaled generators $\bar\chi$ as
\[ \pmatrix{a&b\cr c& d}\ra{\bar a}=\pmatrix{a&b\cr 0&(1-q^{-2})a},\quad
\pmatrix{a&b\cr c& d}\ra{\bar b}=\pmatrix{0&0\cr q^{-2}a&-q^{-2}b}\]
\[ \pmatrix{a&b\cr c& d}\ra{\bar c}=\pmatrix{c&-d+(1-q^{-2})a\cr
0&(1-q^{-2})c},\quad   \pmatrix{a&b\cr c& d}\ra{\bar\xi}=-q^{-2}\pmatrix{a&b\cr
c&d}.\]
As $q\to 1$ this becomes the right action of the super-lie algebra $gl_{1|1}$
on the super-algebra $M_{1|1}$.
\end{example}
\proof The steps are similar to those in the preceding example. This time as
$q\to 1$ one has the even elements $\bar a$ (and $\bar\xi$) acting by matrix
multiplication while
\[\pmatrix{a&b\cr c&d}\ra\bar b=\pmatrix{0&0\cr1&0}\pmatrix{a&b\cr
c&d}\pmatrix{1&0\cr 0&-1},\quad \pmatrix{a&b\cr c&d}\ra\bar c=
\pmatrix{0&1\cr0&0}\pmatrix{a&b\cr c&d}\pmatrix{1&0\cr0&-1}.\]
Note that this is a feature of super-Lie algebras, in the general braided case
(as when $q\ne 1$) even the possibility of a further matrix on the right hand
side will not suffice for a representation as a matrix product. One can verify
directly that these actions represent $gl_{1|1}$ as super-derivations.
\endproof

Thus we recover a complete geometric picture of braided-Lie algebras acting on
braided-commutative algebras of functions (i.e. a classical picture but
braided). The picture unifies the familiar theory of left-invariant vector
fields on groups, super-groups and its obvious generalizations such as to
colour-derivations etc into a single framework based on an $R$-matrix, which
all appear as the semiclassical part of a general braided theory.

\section{Braided Killing Form and the Quadratic Casimir}

In this section we give a final application of our notion of braided-Lie
algebras, namely to the notion of braided-Killing form and associated quadratic
Casimir. It will be $\Ad$-invariant and braided-symmetric in a certain sense.
Like the last section, our a result depends on the fact that we have an actual
finite-dimensional Lie-algebra like subspace $\CL$ or $\CX$ and not merely some
kind of Hopf algebra.

As before, we do the construction first in a categorical setting with diagrams,
and then afterwards deduce and compute the matrix form. The idea behind the
braided Killing form in the categorical setting is quite straightforward. In
any braided category with duals there is a natural notion of braided-trace of
an endomorphism. Assuming that $\CL$ has a dual $\CL^*$ (a kind of
finite-dimensionality condition) we define the braided-Killing form via the
braided-trace in the adjoint representation of $U(\CL)$ on $\CL$ constructed in
Proposition~4.4. We begin with the braided-trace itself.

\begin{propos} For an object $V$ in a braided category with dual $V^*$, and any
morphism $\phi:W\tens V\to V$ we define the {\em braided trace} as the map
$\und\trace(\phi):W\to\und 1$ obtained as shown in Figure~14 (a). If $B$ is a
braided-Hopf algebra and acts cocommutatively by $\alpha$ on $V$ then
$\und\trace(\phi)$ is $B$-invariant in the manner shown in (b).
\end{propos}
\begin{figure}
\vskip .3in
\epsfbox{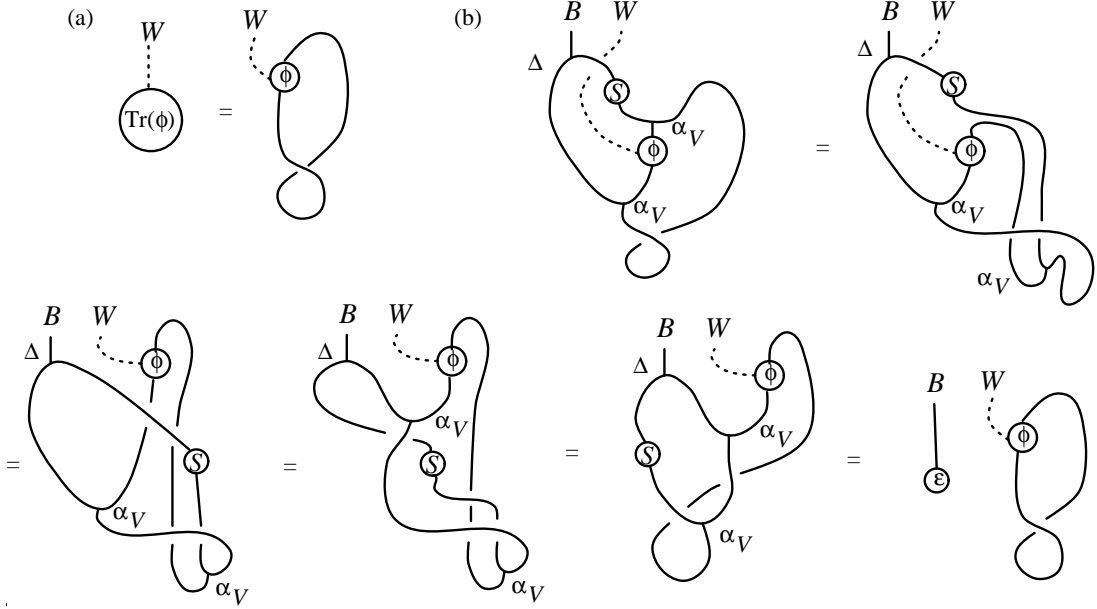}
\caption{Definition (a) of braided-trace $\und\trace$ of a morphism $W\tens
V{\buildrel\phi\over\to}V$ and (b) proof of its cyclicity property of
invariance under a cocommutative action $\alpha$ of any braided-Hopf algebra
$B$. The extra input $W$ is optional}
\end{figure}
\proof By definition $\und\trace(\phi)$ is a morphism $W\to\und 1$ as shown in
(a). Here $\cup$ and $\cap$ denote evaluation $V^*\tens V\to\und 1$ and
coevaluation $\und 1\to V\tens V^*$ respectively. In part (b) we suppose that a
braided-Hopf algebra $B$ acts on $V$ cocommutatively. The first equality uses
functoriality and the double-bend property of duals (compatibility between
evaluation and coevaluation, as used above in Proposition~6.1) to pull
$\alpha_V$ down. The second equality cancels the new double-bend and also
pushes $\phi$ up. The third equality is the braided-cocommutativity of $B$ with
respect to $V$. We then use functoriality to reorganise, and that $\alpha$ is
an action to cancel using the braided-antipode axioms. \endproof

The invariance here is our braided-analog of the usual `cyclicity' property of
the trace. Note also that $W$ can be anything, for example $W=\und 1$ and
$\phi:V\to V$ an endomorphism. We have retained the extra input $W$ for greater
generality. In particular, if $W=B$ and $\phi=\alpha$ then the invariance means
precisely that $\und\trace(\alpha)$ is $Ad$-invariant, where $\alpha$ is the
braided-adjoint action of Section~3.

\begin{propos} Let $\CL$ be a braided-Lie algebra in the setting of Section~4.
We define its braided-Killing form $g:\CL\tens \CL\to \und 1$ to be the
braided-trace of the map $[\ , \ ]\circ(\id\tens[\ ,\ ])$. In concrete terms
this is
\[ g(\xi,\eta)=\und\trace([\xi,[\eta,\ ]])\]
for $\xi,\eta\in \CL$. If $U(\CL)$ has an antipode then  $g$ is invariant under
$[\ ,\ ]$ as shown in Figure~15 (c). It is braided-symmetric as shown in
Figure~15 (d). The braided-Killing form is defined on all of $U(\CL)\tens
U(\CL)$ and has descendants $T$ and $\und\dim(\CL)$ as also shown.
\end{propos}
\begin{figure}
\vskip .3in
\epsfbox{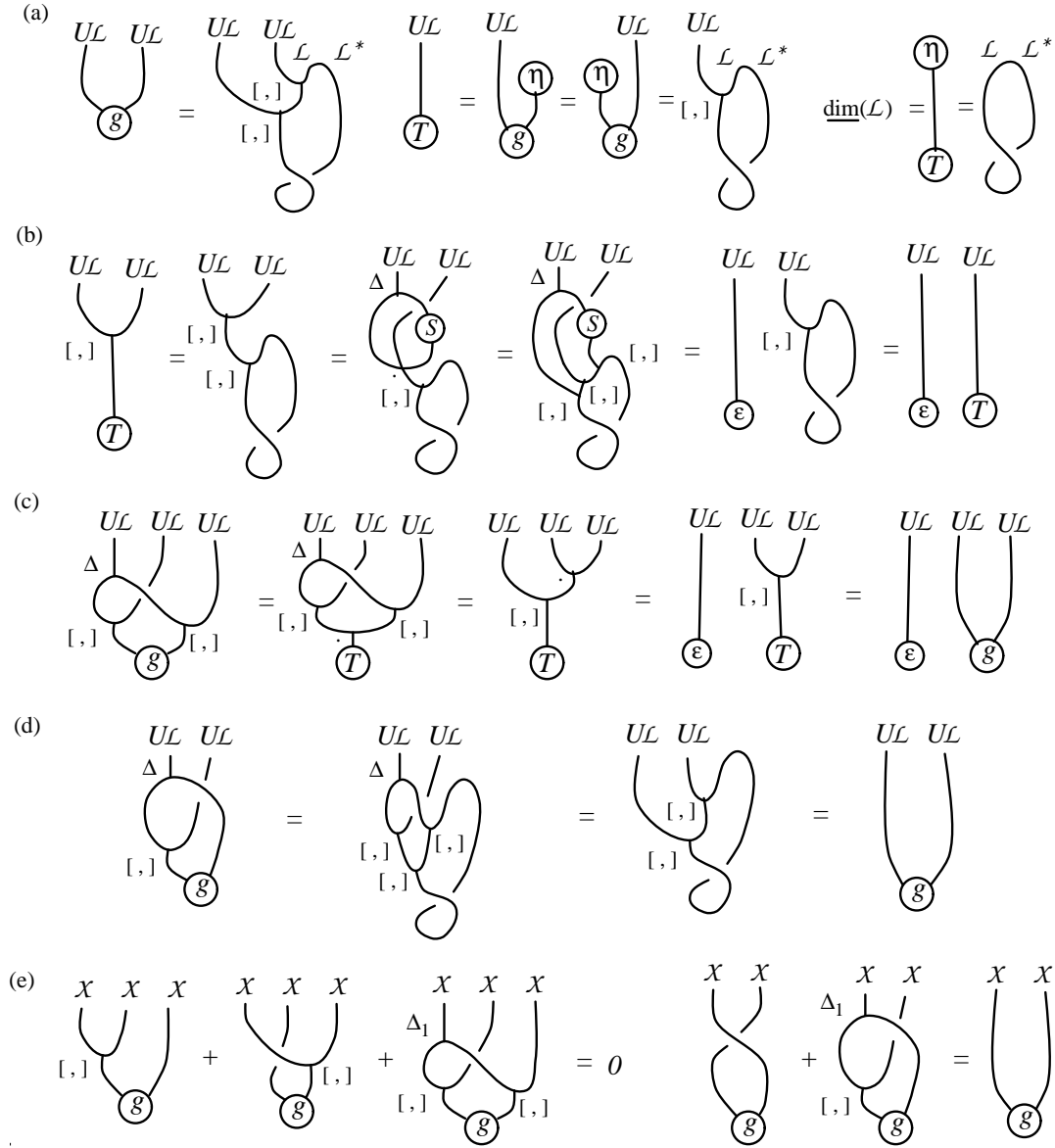}
\caption{Definition (a) of braided-Killing form and its descendants (b)-(c)
proof of their $[\ ,\ ]$-invariance and (d) braided-symmetry. In the form (e)
on $\CX$ these look more familiar.}
\end{figure}
\proof The braided-metric is defined as the braided-trace of the iterated
braided-adjoint action. This is well-defined as a morphism $\CL\tens\CL\to\und
1$ but can also be viewed as shown in (a) as the restriction of a morphism
$U(\CL)\tens U(\CL)\to \und 1$. In this case, because $[\ ,\ ]$ is an action,
we can understand it as multiplication in $U(\CL)$ followed by the
braided-trace in the braided-adjoint representation. In this case its
$Ad$-invariance follows at once in (c) from the $Ad$-invariance of $T$ proven
in part (b). This in turn follows from the cyclicity of the braided-trace
proven in Proposition~7.1. This assumes in the second equality that $U(\CL)$
has a braided-antipode, in which case $[\ ,\ ]$ can be identified with the
braided-adjoint action as explained in Section~4. Part (d) is the
braided-symmetry property. The first equality is the definition of $g$, the
second is the extended-form of the braided-Jacobi identity in Section~4. For
the braided-symmetry only on $\CL$ we need only the braided-Jacobi identity
axiom (L1). Finally, part (e) justifies our terminology by showing how the
property looks on the subspace $\CX\subset U(\CL)$ where the coproduct is as in
Figure~11. \endproof

Clearly the braided-Killing form is the same as first multiplying in $U(\CL)$
and then applying the braided-trace to $[\ ,\ ]$ considered as an action of
$U(\CL)$ from Proposition~4.4. Also, if $\CL$ is of the form $\CL_1=1\oplus\CX$
as discussed at the end of Section~4, we can equally well define
\[ g_{\chi}:\CX\tens \CX\to \und 1\]
in just the same way as $\und\trace([\ ,\ ](\id\tens[\ ,\ ])$ restricted to
$\chi\tens\chi$. Both are useful in examples. The metric on $\CL$ is some kind
of `multiplicative' Killing form while $g_{\chi}$ is more like the classical
one. Its diagrammatic properties are in Figure~15(e).

The proof above assumes that $U(\CL)$ has a braided-antipode. On the other hand
the formulation of the proposition does not require this if we work with $[\ ,\
]$ instead of an actual braided-adjoint action. This was the strategy in
Section~4 and we take the same view here. For example, in the tensor setting of
Proposition~5.1 we can assume that the tensors defining the braided-Lie algebra
are sufficiently nice for $\CL$ to have a dual object and for the
braided-Killing form to be $[\ ,\ ]$-invariant. We say in this case that the
braided-Lie algebra is {\em regular}. Also, we define tensors for $g$ and the
braided trace, as well as the normalization $\und\dim(\CL)$ by
\eqn{met-tens}{g(u^I\tens u^J)=\und\trace([u^I,[u^J,\ ]])=g^{IJ},\quad
\und\trace([u^I,\ ])=T^I,\quad \und\dim(\CL)=\und\trace(\id)}
Their properties in tensor form are read of from the braid-diagrams just as for
Proposition~5.1. In particular, the invariance and braided-symmetry conditions
take the form
\eqn{tens-g-inv}{\Delta^I{}_{AB}\, R^J{}_M{}^B{}_N\, c^{AM}{}_P\, c^{NK}{}_Q
g^{PQ}=\delta^I\, g^{JK},\quad c^{IJ}{}_K T^K=\delta^I T^J}
\eqn{tens-g-sym}{\Delta^I{}_{AB}\, R^J{}_M{}^B{}_N\, c^{AM}{}_P\,
g^{PN}=g^{IJ}}
and likewise for $g_{\chi}$ and $T_\chi$ on the generators
$\chi^I=u^I-\delta^I$. These are related to $g^{IJ}$ and $T^I$ by
\eqn{tens-g-chi}{ g_{\chi}^{IJ}=g(\chi^I\tens\chi^J)=g^{IJ}-\delta^I
T^J-\delta^J T^I+\und\dim(\CL)\delta^I\delta^J,\quad
T_\chi^I=T^I-\und\dim(\CL)\delta^I.}
Here $g$ and $g_{\chi}$ differ only by the braided-trace of the action of $1$
in one or other or both of the inputs.
The fact that these maps are all morphisms in the category means that they obey
the corresponding morphism conditions along the lines of (L0) in
Proposition~5.1. Thus, $T^J$ obeys the same equations as for $\delta^I$ in (L0)
while
$g$ (and $g_\chi$) obey
\eqn{g-morph}{R^{K}{}_M{}^J{}_B\, R^M{}_L{}^I{}_A\,
g^{AB}=g^{IJ}\delta^{K}{}_L,\quad R^I{}_A{}^K{}_M\, R^J{}_B{}^M{}_L\,
g^{AB}=g^{IJ}\delta^{K}{}_L.}

We have mentioned in the proof of Proposition~5.1 that the nicest setting is
the one in which the constructions can be viewed as taking place in the
category of left $A({\bf R})$-comodules, or more precisely in the category of
$A$-comodules where $A$ is a dual-quasitriangular quotient of $A({\bf R})$. In
the present context one could demand also that $A$ is
a Hopf algebra. In this case its category of comodules has duals, so this is
sufficient to have a quantum trace. We do not want to limit ourselves to this
case, but it is convenient for generating the necessary formulae which can then
be verified directly on the assumption of suitable properties for the structure
constants. To see that this supposition implies restrictions on $A$ we note
that in these terms, the morphism properties of $\und\Delta,\und\eps,c,g$ are
\eqn{A-cov.lie}{ t^I{}_J\Delta^J{}_{KL}=\Delta^I{}_{AB} t^A{}_K t^B{}_L,\quad
t^I{}_J\delta^J=\delta^I,\quad  c^{IJ}{}_K t^K{}_L=t^I{}_A t^J{}_B c^{AB}{}_L}
\eqn{A-cov.g.t}{g^{IJ}=t^I{}_A t^J{}_B g^{AB},\quad t^I{}_JT^J=T^I}
where $t^I{}_J$ is the matrix generator of $A({\bf R})$.

\begin{propos} Let $\CL$ be a braided-Lie algebra of the general tensor type in
Proposition~5.1 and suppose that it lives in the category of $A$-comodules as
explained. Then
\[ g^{IJ}= c^{IA}{}_B\, c^{JL}{}_A\, \widetilde{R}{}^K{}_L{}^B{}_K, \quad T^I=
c^{IJ}{}_A\, \widetilde{R}{}^K{}_J{}^A{}_K,\quad
\und\dim(\CL)=\widetilde{R}{}^K{}_J{}^J{}_K\]
where  $\widetilde{\ }$ denotes the second-inverse as above but applied now to
the multi-index $\bf R$.
\end{propos}
\proof We assume here that the category in which we work is the braided tensor
category of left $A$-comodules where $A$ is a dual-quasitriangular Hopf algebra
given as a quotient of $A({\bf R})$. It has at least the additional relations
(\ref{A-cov.lie}) and (\ref{A-cov.g.t}) as explained. The finite-dimensional
comodules such as $\CL$ and $\CX$ here then have duals in the category using
the antipode. From this one computes the braiding between a basis $\{u^I\}$ of
$\CL$ and a dual basis $\{f_I\}$ say of $\CL^*$ in a standard way as explained
in \cite{Ma:lin}. The $\{u^I\}$ transform as a vector under the matrix
generator of $A({\bf R})$ and $\{f_I\}$ as a covector with right-multiplication
by the inverse matrix generator.
This gives
\[ \Psi(u^I\tens f_J)=f_K\tens u^L\widetilde{R}{}^K{}_J{}^I{}_L.\]
Using this for the braid-crossing in the diagrammatic definition of the
braided-trace and braided-Killing form and proceeding as in Proposition~5.1 for
the other tensors, immediately gives the results stated. Note that the
$\und\trace$ that we use here is defined for any endomorphism $\phi^I{}_J$ by
$\und\trace(\phi)=\phi^B{}_A\widetilde{R}{}^K{}_B{}^A{}_K$ just as for the
usual quantum or braided trace associated to an $R$-matrix. We are simply using
this now applied to the endomorphisms built from the structure constants
$c^{IJ}{}_K$ of the braided-Lie algebra.
\endproof

In our matrix examples of Proposition~5.2, all the data are based on an initial
$R$-matrix $R^i{}_j{}^k{}_l$. In this context we have already introduced the
notion for quantum groups that $R$ is regular if $A(R)$ has a quotient Hopf
algebra $A$ which remains dual-quasitriangular. In this case $B(R)$ has a
quotient which is indeed a braided-Hopf algebra with braided-antipode. Related
to this, $U(\CL)=B(R)$ for this class of matrix-braided-Lie algebras is indeed
regular in the sense above. On the other hand, we do not want to limit
ourselves to this case. In fact, it is sufficient to suppose that $R$ obeys
certain matrix identities to arrive at the same conclusion.

\begin{propos}
In our matrix examples of Proposition~5.2 we suppose that this is regular in
the sense that the initial $R\in M_n\tens M_n$ comes from a quantum group
obtained from $A(R)$. Then the braided-Killing  $g$ is given by
\[ g^{IJ}=c^{IA}{}_B\, c^{JL}{}_A\, R^{b_0}{}_{c}{}^{a}{}_{n}
R{}^{n}{}_{d}{}^{c}{}_{b}  \vartheta^b{}_{l_0}
\tilde{R}^{d}{}_{b_1}{}^{l_1}{}_{a}\]
in terms of the initial $R$ and its second-inverse $\tilde{R}$. Here
$\vartheta^i{}_j=\widetilde{R}{}^i{}_k{}^k{}_j$. Similarly for $T^I$ and
$\und\dim(\CL)$. If $R=R_0+O(\hbar)$ then $g_{\chi}=O(\hbar^2)$. On the
rescaled generators $\bar\chi^I=\hbar^{-1}\chi^I$ we have
\[ g_{\bar\chi}^{IJ}=g(\bar\chi^I,\bar\chi^J)=K^{IJ}+O(\hbar),\quad
T_{\bar\chi}^I=T(\bar\chi^I)=O(\hbar)\]
where $K^{IJ}$ defines the Killing form of the $R_0$-Lie algebra in
Proposition~5.3  and has its usual Ad-invariance and $\Psi$-symmetry properties
(e.g. for usual, super or colour Lie algebras etc). Here $\Psi=\Psi(R_0)$ is
symmetric. Meanwhile, the braided trace $T$ on the rescaled generators tends to
zero.
\end{propos}
\proof One can either compute $\widetilde{R}{}^K{}_J{}^A{}_K$ for the
particular matrix in Proposition~5.2, or compute the braiding $\Psi(u^I\tens
f_J)$ between a basis element of $u^I$ and a dual-basis element directly in the
same way that the braiding in Proposition~5.2 was obtained in
\cite{Ma:eul}\cite{Ma:exa}. For the latter course the category in which we work
is that of right $A$-comodules where $A$ is now a dual-quasitriangular Hopf
algebra obtained as a quotient of $A(R)$ and $R^i{}_j{}^k{}_l$ here is the
initial $R$-matrix in in Proposition~5.1. It is related to the general setting
above via the bialgebra map $A({\bf R})\to A^{\rm op}$ given by $t^I{}_J\mapsto
 t^{i_0}{}_{j_0}\cdot_{A^{\rm op}}S_{A^{\rm op}}t^{j1}{}_{i1}$. This along with
the antipode of $A$ converts the left-comodule algebras in the general setting
into right $A$-comodule algebras. In the latter category the elements $\vecu$
transform under the right adjoint coaction $\vecu\to\vect^{-1}\vecu\vect$ using
a compact notation where $\vect$ is the matrix generator of $A(R)$. This
induces on the dual basis $\{f_i{}^j\}$ the transformation $f_i{}^j\to f_m{}^n
\tens (St^j{}_n)S^2 t^m{}_i$ where $S$ is the antipode. From this one has
\align{\Psi(u^{i_0}{}_{i_1}\tens f_{j_0}{}^{j_1})&&=f_{k_0}{}^{k_1}\tens
u^{l_0}{}_{l_1}\CR((St^{i_0}{}_{l_0})t^{l_1}{}_{i_1}\tens
(St^{j_1}{}_{k_1})S^2t^{k_0}{}_{j_0})\\
&&=f_{k_0}{}^{k_1}\tens u^{l_0}{}_{l_1}\CR(St^{i_0}{}_{l_0}\tens
(St^{a}{}_{k_1})S^2t^{k_0}{}_{b})\CR(t^{l_1}{}_{i_1}\tens
(St^{j_1}{}_a)S^2t^b{}_{j_0})\\
&&=f_{k_0}{}^{k_1}\tens u^{l_0}{}_{l_1}
\CR(t^{i_0}{}_{c}\tens t^{a}{}_{k_1})
\CR(t^{c}{}_{l_0}\tens S t^{k_0}{}_{b})
\CR(t^{l_1}{}_{d}\tens S^2t^{b}{}_{j_0})
\CR(t^{d}{}_{i_1}\tens St^{j_1}{}_{a})\\
&&=f_{k_0}{}^{k_1}\tens u^{l_0}{}_{l_1}
R^{i_0}{}_{c}{}^{a}{}_{k_1}
\tilde{R}^{c}{}_{l_0}{}^{k_0}{}_{b}
(\tilde R)^{-1}{}^{l_1}{}_{d}{}^{b}{}_{j_0}
\tilde{R}^{d}{}_{i_1}{}^{j_1}{}_{a}=f_K\tens u^L\widetilde{R}{}^K{}_J{}^I{}_L}
where in the last line we evaluated the dual-quasitriangular structure $\CR$ on
the matrix generators. This gives the matrix $\widetilde{R}{}^K{}_J{}^I{}_L$ in
this example (compare with the braiding in the proof of the previous
proposition). Composing this with evaluation we have
\align{<\ ,\ >\circ\Psi(u^I\tens f_J)&&=\widetilde{R}{}^K{}_J{}^I{}_K=
R^{i_0}{}_{c}{}^{a}{}_{n}\tilde{R}^{c}{}_{m}{}^{m}{}_{b}
(\tilde R)^{-1}{}^{n}{}_{d}{}^{b}{}_{j_0}
\tilde{R}^{d}{}_{i_1}{}^{j_1}{}_{a}\\
&&= R^{i_0}{}_{c}{}^{a}{}_{n} \vartheta^c{}_b
(\tilde R)^{-1}{}^{n}{}_{d}{}^{b}{}_{j_0}
\tilde{R}^{d}{}_{i_1}{}^{j_1}{}_{a}=R^{i_0}{}_{c}{}^{a}{}_{n}
R{}^{n}{}_{d}{}^{c}{}_{b}  \vartheta^b{}_{j_0}
\tilde{R}^{d}{}_{i_1}{}^{j_1}{}_{a}}
where $\vartheta^c{}_b=\tilde{R}^{c}{}_{m}{}^{m}{}_{b}$ is the matrix used for
the quantum or braided trace associated to the initial $R$-matrix. It obeys
$\vartheta_2(\tilde{R})_{12}^{-1}\vartheta_2^{-1}=R_{12}$ as proven in
\cite{Ma:lin} and we use this now. Putting this into the preceding proposition
gives the results stated.
Note that more generally, one can suppose only that $R$ is bi-invertible and
obeys suitable matrix identities such as\cite{Ma:lin}.
\eqn{varthetaid}{\vartheta_2\widetilde{R}=R^{-1}\vartheta_2,\quad
\vartheta_1R^{-1}=\widetilde{R}\vartheta_1}
to conclude the proposition directly.

{}From this one sees the limit as $R$ approaches a triangular solution $R_0$.
{}From Figure~15~(e) we see that the semiclassical part $K$ of the
braided-Killing has the familiar properties. Likewise for the braided-trace.
\endproof

Thus the braided-Killing form reduces near the triangular solutions to the more
usual notion of Killing form which is $Ad$-invariant and $\Psi$-symmetric in
the more naive sense. This includes of course the usual Killing form but holds
also for super-Lie algebras and colour-Lie algebras. In the latter cases we
have not found this notion in the literature, perhaps because it need not be
non-degenerate as we shall see in an example. In the former standard case we
will recover the usual Killing form which will be non-degenerate on the
semisimple part of the classical limit.
We find here an unusual phenomenon: the process of $q$-deformation can make a
degenerate Killing form non-degenerate.

\begin{example} In Example~5.5 where $R=R_{gl_2}$ the braided-Killing form and
trace etc on $\CL$ is
\[ g={[4]_q\over q^{10}}\pmatrix{q^4+q^{-2}-1&0&0&q^4-q^2+1\cr
0&0&(1-q^{-2})^2&0\cr
0&(q-q^{-1})^2&0&0\cr
q^4-q^2+1&0&0& q^4-q^2+1+(1-q^{-2})^2}\]
\[ T^I=(1+[3]_{q^2})\delta^I,\quad \und\dim(\CL)=[4]_{q};\qquad
[n]_q={1-q^{-2n}\over 1-q^{-2}}.\]
Here $g$ is non-degenerate for generic $q$. The braided-Killing form on the
rescaled $\bar{\chi}$ with basis $\bar\xi,\bar b,\bar c,\bar\gamma$ is also
non-degenerate for generic $q$ and given by
\[ g_{\bar\chi}=q^{-4}\pmatrix{[4]_q[2]_q&0&0&0\cr
0&0&q^{-4}[4]_q&0\cr
0&q^{-2}[4]_q&0&0\cr
0&0&0&[3]_q[2]_q(1-q^{-4})}\]
As $q\to 1$ it becomes the usual Killing form on $sl_2$ and $0$ on the
$\bar\gamma$ generator.
\end{example}
\proof This is a direct computation from Proposition~7.2 or 7.3 (the result it
the same). Note that as $q\to 1$ the
braided-Killing forms become symmetric and the braided-traces of the
$\bar\chi^I$ become zero as we should expect from Proposition~7.4. \endproof

It is remarkable here that the braided-Killing form on our braided-version of
$gl_2$ is non-degenerate for generic $q$. This reflects the fact that for
generic $q$ the $U(1)$ generator $\bar\gamma$ in Example~5.5 did not fully
decouple from the braided-Lie bracket. This is in spite of the fact that it is
central in the braided-enveloping algebra.

\begin{example} In Example~5.6 where $R=R_{gl_{1|1}}$ the braided-Killing form
on $\CL$ and on the $\bar\chi$ $\bar a,\bar b,\bar c,\bar\xi$ basis are
\[g=-(q^2-q^{-2})^2\pmatrix{1&0&0&1\cr
0&0&0&0\cr 0&0&0&0\cr 1&0&0& 1},
\quad g_{\bar\chi}=-(1+q^{-4})\pmatrix{1&0&0&0\cr 0&0&0&0\cr0&0&0&0\cr0&0&0&0}
 \]
\[ T^I=-(q-q^{-1})^2\delta^I,\quad \und\dim(\CL)=0\]
\end{example}
\proof This is likewise a direct computation from Proposition~7.3 or 7.4.
\endproof

The braided-dimension becomes as $q\to 1$ the super-dimension for the
$R$-matrix in this example. Hence its vanishing corresponds in the limit to the
equal number of bose and fermi modes in the algebra. This is typical of
vanishing theorems in super-symmetry and suggests that similar results can
sometimes extend to the braided case. A similar degeneracy of the
braided-Killing form, and vanishing of the braided dimension holds for other
non-standard $R$-matrices (such as the 8-vertex model solution). On the other
hand non-degeneracy as in Example~7.5 is typical of the standard $R$-matrices
associated to deformations of semisimple Lie algebras.

Armed with non-degeneracy in at least some cases it is natural to define for
invertible $g^{IJ},g_{\chi}^{IJ}$ the corresponding quadratic Casimirs in
$U(\CL)$,
\eqn{casimirs}{C=u^Iu^Jg_{IJ},\quad C_\chi=\chi^I\chi^Jg_\chi{}_{IJ}}
where the matrixes with lower indices are the matrix inverses. This can also be
said diagrammatically.

\begin{corol} In the setting of Proposition~7.2 we suppose that the
braided-Killing form has an inverse $g:\und 1\to \CL\tens \CL$. Then this is
$[\ ,\ ]$-invariant and the Casimir $\cdot\circ g:\und 1\to U(\CL)$ is
invariant and central in $U(\CL)$. Moreover, the braided-Killing form and its
inverse allow us to identify $\CL$ and $\CL^*$ in the category.
\end{corol}
\begin{figure}
\vskip .3in
\epsfbox{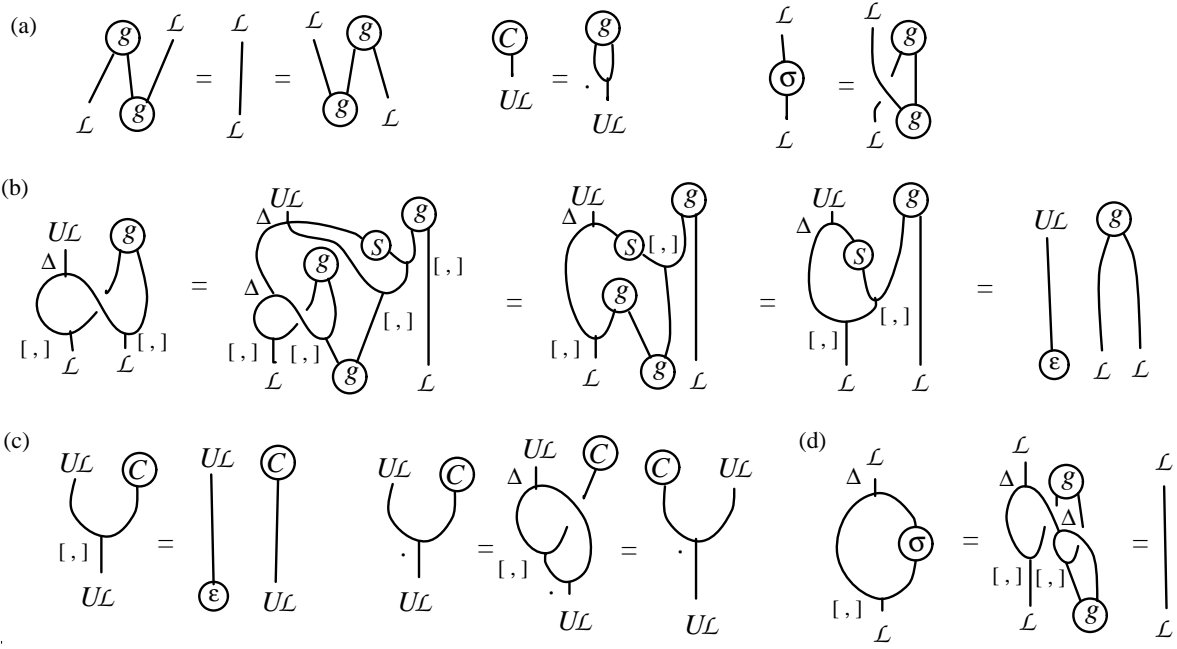}
\caption{Definition (a) inverse braided Killing form, braided-quadratic Casimir
and associated twist morphism. $[\ ,\ ]$-invariance (b) of $g$ implies
invariance and centrality (c) of the braided-Casimir. (d) is a property of the
twist $\sigma$ related to braided-antisymmetry of the bracket.}
\end{figure}
\proof The categorical inverse (also denoted $g$) is defined via Figure~16 (a),
along with some related maps. The corollary then follows at once from the
invariance and braided-symmetry properties of the braided-Killing form in
Figure~15, as shown. For simplicity we have assumed for the proof that $U(\CL)$
has a braided-antipode, but as in Proposition~7.2 we do not limit ourselves to
this case. In the matrix example of Proposition~5.2 with suitable $R$ one can
prove the invariance of the inverse-Killing form directly. Also, using these
maps we can identify $\CL$ with $\CL^*$,  with the braided-Killing form as
evaluation and its inverse as co-evaluation. In this case it is natural to
consider the twist $\sigma$ and we have included in part (d) one of its
interesting properties. \endproof

 Similar properties (with similar proofs) apply to the inverse $g_\chi:\und
1\to \CX\tens\CX$ when this exists. The identification of $\CL$ with $\CL^*$
(or $\CX$ with ${\CX}^*$) when the inverses exist means in tensorial terms that
we can use the braided-metric and its inverse to raise and lower indices in a
familiar way.

\begin{example} For the braided-Lie algebra in Example~5.5 the quadratic
Casimirs defined from the inverse of the braided-Killing form are central and
take the form
\[ C={[2]_q\over [4]_q(1-q^{-2})^2}\left({(q^6-q^2+1)q^4\over
(1+q^2)(q^8+q^4-q^2+1)}(q^{-2}a+d)^2-(ad-q^2cb)\right)\]
\[ C_\chi={q^4\over [4]_q}({\bar\xi^2\over [2]_q}+\bar b\bar c+q^2\bar c\bar
b)+{q^2\over [3]_q(1-q^{-4})^2}\bar\gamma^2\]
As $q\to 1$ the $sl_2$-part of $C_\chi$ tends to the usual quadratic Casimir
and the $U(1)$ part tends to $\infty$.
\end{example}
\proof This is by direct computation from the generators using REDUCE. To put
the results into the form shown we made extensive use of the relations in
$U(\CL)=BM_q(2)$ from \cite{Ma:exa}. We know that the rescaled generators tend
in the classical limit to $gl_2$. We see that the natural braided Casimir tends
to the usual quadratic Casimir for the $sl_2$ part while the $U(1)$ part blows
up in terms of the rescaled generator $\bar\gamma$. Note that the two terms in
$C_\chi$ are separately central for all $q$ so one an subtract off this
divergent part if desired. \endproof

Moreover, for $q\ne 1$ and the standard $R$-matrices we can put here the form
$\vecu=l^+Sl^-$ and recover from $C$ the (square of the) quantum quadratic
Casimir known previously by other methods. On the other hand our construction
is not tied to such standard cases.

This completes our development of the basic theory of braided-Lie algebras and
some typical examples. Further applications and examples will be developed
elsewhere. The phenomenon seen here for the braided version of $gl_2$ can be
expected quite generally and is part of one set of potential applications of
the theory, namely to a process that can be called `q-regularization' of
singularities. The singularity of the inverse-Killing form for $gl_2$ is
resolved by $q$-deformation in our braided context, as a pole at $q=1$. The
regularization of infinities in physics is one of the motivations for
$q$-deformed physics and $q$-deformed geometry (another is interesting
phenomena at roots of unity).

For one possible physical application of these constructions we note that we
have introduced a general quantum-group gauge theory in \cite{BrzMa:gau}, which
should adapt (by transmutation) to our braided-setting. The gauge fields of
such a theory should take values in a braided-Lie algebra $\CL$ or $\CX$ and
the Yang-Mills Lagrangian should involve the braided-Killing form as above. In
such a theory the $SU(2)\times U(1)$ of the standard model could be unified for
$q\ne 1$ with the $U(1)$ mode not decoupling from the $SU(2)$ mode in the bare
Lagrangian. After renormalization the zero in the braided-Killing form above
for the $U(1)$ mode may still leave a residue as the $q$-regularization is
removed.

For another direction we note that the quadratic Casimirs become represented as
differential operators on the braided-group function algebras as we have seen
in Section~6. Thus
\eqn{laplacian}{
\overleftarrow{\lform}=\overleftarrow{\del^I}\overleftarrow{\del^J}g_{IJ},\quad
\overleftarrow{\lform}{}_\chi=\overleftarrow{D^I}
\overleftarrow{D^J}g_{\chi}{}_{IJ}}
should play the role of Laplacian in some kind of braided-geometry of which the
braided-groups are the simplest examples. They could perhaps be used as
propagators in some form of braided or q-deformed physics. Again, one would
have in mind some interaction with the process of renormalization, where $q$ is
regarded as a regularization parameter and set to 1 at the end. It may also be
that $q\ne 1$ could be used as a model of feedback to the geometry due to
quantum effects in the context of Planck scale physics.

Related to these considerations we note that there are further examples of
braided Hopf algebras associated to the quantum plane and to the
braided-Heisenberg algebras\cite{Ma:fre}, as well as possibly to the
infinite-dimensional exchange algebras in conformal field theory -- one would
like to know if they have a braided-Lie algebra underlying them. The
deformation of braided-Lie algebras via braided-Poisson brackets is a further
question related to these.

Apart from these physical directions, there are of a variety of natural
mathematical questions also to be addressed. The long term goal is to develop
the differential geometry of braided groups with braided-Lie algebras and other
braided-geometrical constructions in analogy with the classical theory. In this
paper we have taken some of the first  steps in such a programme. We introduced
brackets, vector-field or matrix realizations and Killing forms for them in
some generality. We recover usual notions from any regular $R$-matrix, which
need not be a standard deformation of the identity. The theory interpolates and
unifies with super and other Lie-algebra constructions also. Moreover, even in
the standard case we have found some unusual phenomena concerning the removal
of degeneracy.

\end{document}